\documentclass[fleqn,usenatbib]{mnras}

\usepackage{newtxtext,newtxmath}

\usepackage[T1]{fontenc}
\usepackage{ae,aecompl}

\usepackage{comment}
\usepackage{graphicx}	
\usepackage{amsmath}	
\usepackage{amssymb}	
\usepackage{multirow}
\usepackage[usestackEOL]{stackengine}
\usepackage{multicol}        %
\usepackage{bm}		%
\usepackage{pdflscape}	%
\usepackage[left]{lineno}
\usepackage{fontawesome}
\usepackage{xcolor}

\newcommand{\lkhd}{\ensuremath{\mathcal{L}}}
\newcommand{\photoz}{\ensuremath{\mathrm{photoz}}}

\title[$\mu_{\star}$ mass-calibration on DES Y1]{$\mu_{\star}$ Masses: Weak Lensing Calibration of the  Dark Energy Survey Year 1 redMaPPer Clusters using Stellar Masses}

\author[M. E. S. Pereira et al.]{
\parbox{\textwidth}{
\Large
M.~E.~S.~Pereira,$^{1}$\thanks{Contact e-mail: \href{mailto:mariaeli@brandeis.edu}{mariaeli@brandeis.edu}}
A.~Palmese,$^{2,3}$
T.~N.~Varga,$^{4,5}$
T.~McClintock,$^{6}$
M.~Soares-Santos,$^{1}$
J.~Burgad,$^{7}$
J.~Annis,$^{2}$
A.~Farahi,$^{8}$
H.~Lin,$^{2}$
A.~Choi,$^{9}$
J.~DeRose,$^{10,11}$
J.~Esteves,$^{1}$
M.~Gatti,$^{12}$
D.~Gruen,$^{13,14,15}$
W.~G.~Hartley,$^{16,17,18}$
B.~Hoyle,$^{4,5}$
T.~Jeltema,$^{11}$
N.~MacCrann,$^{9,19}$
A.~Roodman,$^{14,15}$
C.~S{\'a}nchez,$^{20}$
T.~Shin,$^{20}$
A.~von der Linden,$^{21}$
J.~Zuntz,$^{22}$
T.~M.~C.~Abbott,$^{23}$
M.~Aguena,$^{24,25}$
S.~Avila,$^{26}$
E.~Bertin,$^{27,28}$
S.~Bhargava,$^{29}$
S.~L.~Bridle,$^{30}$
D.~Brooks,$^{17}$
D.~L.~Burke,$^{14,15}$
A.~Carnero~Rosell,$^{31,32}$
M.~Carrasco~Kind,$^{33,34}$
J.~Carretero,$^{12}$
M.~Costanzi,$^{35,36}$
L.~N.~da Costa,$^{25,37}$
S.~Desai,$^{38}$
H.~T.~Diehl,$^{2}$
J.~P.~Dietrich,$^{39}$
P.~Doel,$^{17}$
J.~Estrada,$^{2}$
S.~Everett,$^{11}$
B.~Flaugher,$^{2}$
P.~Fosalba,$^{40,41}$
J.~Frieman,$^{2,3}$
J.~Garc\'ia-Bellido,$^{26}$
E.~Gaztanaga,$^{40,41}$
D.~W.~Gerdes,$^{42,8}$
R.~A.~Gruendl,$^{33,34}$
J.~Gschwend,$^{25,37}$
G.~Gutierrez,$^{2}$
S.~R.~Hinton,$^{43}$
D.~L.~Hollowood,$^{11}$
K.~Honscheid,$^{9,19}$
D.~J.~James,$^{44}$
K.~Kuehn,$^{45,46}$
N.~Kuropatkin,$^{2}$
O.~Lahav,$^{17}$
M.~Lima,$^{24,25}$
M.~A.~G.~Maia,$^{25,37}$
M.~March,$^{20}$
J.~L.~Marshall,$^{47}$
P.~Melchior,$^{48}$
F.~Menanteau,$^{33,34}$
R.~Miquel,$^{49,12}$
R.~L.~C.~Ogando,$^{25,37}$
F.~Paz-Chinch\'{o}n,$^{50,34}$
A.~A.~Plazas,$^{48}$
A.~K.~Romer,$^{29}$
E.~Sanchez,$^{51}$
V.~Scarpine,$^{2}$
M.~Schubnell,$^{8}$
S.~Serrano,$^{40,41}$
I.~Sevilla-Noarbe,$^{51}$
M.~Smith,$^{52}$
E.~Suchyta,$^{53}$
M.~E.~C.~Swanson,$^{34}$
G.~Tarle,$^{8}$
R.~H.~Wechsler,$^{13,14,15}$
J.~Weller,$^{4,5}$
and Y.~Zhang$^{2}$
\vspace{0.3cm}
\begin{center} (DES Collaboration) \end{center}
}\vspace{0.3cm} 
\\
{\small\emph{(Affiliations are listed at the end of paper)} }
}

\date{Accepted XXX. Received YYY; in original form ZZZ}

\pubyear{2020}

\begin{document}
\label{firstpage}
\pagerange{\pageref{firstpage}--\pageref{lastpage}}
\maketitle

\begin{abstract}
 
We present the weak lensing mass calibration of the stellar mass based $\mu_{\star}$ mass proxy for redMaPPer galaxy clusters in the Dark Energy Survey Year 1. For the first time we are able to perform a calibration of $\mu_{\star}$ at high redshifts, $z>0.33$. In a blinded analysis, we use $\sim 6,000$ clusters split into 12 subsets spanning the ranges $0.1 \leqslant z<0.65$ and $\mu_{\star}$ up to $\sim 5.5 \times 10^{13} M_{\odot}$, and infer the average masses of these subsets through modelling of their stacked weak lensing signal. In our model we account for the following sources of systematic uncertainty: shear measurement and photometric redshift errors, miscentring, cluster-member contamination of the source sample, deviations from the NFW halo profile, halo triaxiality and projection effects. We use the inferred masses to estimate the joint mass--$\mu_{\star}$--$z$ scaling relation given by $\langle M_{200c} | \mu_{\star},z \rangle = M_0 (\mu_{\star}/5.16\times 10^{12} \mathrm{M_{\odot}})^{F_{\mu_{\star}}} ((1+z)/1.35)^{G_z}$. We find $M_0= (1.14 \pm 0.07) \times 10^{14} \mathrm{M_{\odot}}$ with $F_{\mu_{\star}}= 0.76 \pm 0.06$ and $G_z= -1.14 \pm 0.37$. We discuss the use of $\mu_{\star}$ as a complementary mass proxy to the well-studied richness $\lambda$ for: $i)$ exploring the regimes of low $z$, $\lambda<20$ and high $\lambda$, $z \sim 1$; $ii)$ testing systematics such as projection effects for applications in cluster cosmology.       

\end{abstract}




\section{Introduction}

Galaxy clusters are an important tool for studying the formation and evolution of structure in the Universe, the distribution of matter, and for testing modified gravity models. The count of galaxy clusters as a function of mass and redshift is potentially one of the most powerful cosmological probes \citep{2001ApJ...553..545H,RevModPhys.77.207,2011ARA&A..49..409A,2012ARA&A..50..353K,2013SSRv..177....1E,2014JCAP...05..039P,2015Sci...347.1462H,2016arXiv160407626D}. In order to achieve this potential, it is necessary to understand and correct for the systematics involved in the cluster mass calibration, which is currently the dominating source of uncertainties for using clusters to probe cosmology \citep{2010ApJ...708..645R,2015MNRAS.446.2205M, 2016A&A...594A..24P, 2019MNRAS.488.4779C, 2019PASJ...71..107M, 2020arXiv200211124D}. 

Galaxy clusters act as powerful gravitational lenses because their large gravitational fields produce distortions in the shape of the background galaxies. This effect does not depend on the dynamical state of the cluster (as does the X-ray luminosity) and is sensitive to all of its matter content (both baryonic and dark matter). Therefore, using this effect we can assess their matter content and perform very precise mass measurements. Only the most massive clusters have weak lensing signals sufficiently strong to be individually measured in the current generation of wide-field surveys. Thus, we combine the lensing signal of a large number of clusters with similar properties (i.e. \textit{stacking}) to obtain measurements of higher signal-to-noise. 

Observationally, we cannot assess the true mass of galaxy clusters, but we can rank them by some proxy for mass. A mass-observable relation (MOR) must be calibrated to connect the observable and the halo mass. The technique of stacking the weak lensing signal of many systems in a given observable interval provides one of the most direct and model independent methods to calibrate the MORs. The community has made a concerted effort to determine the scaling relations empirically \citep{2001ApJ...554..881S, 2007arXiv0709.1159J, 2014MNRAS.439....2V, 2014MNRAS.443.1973V, 2014MNRAS.439...48A, 2014MNRAS.444..147O, 2015MNRAS.449..685H, 2015MNRAS.447.1304F, 2015ApJ...807..178W, 2015MNRAS.452..701W, 2015MNRAS.446.2205M, 2016MNRAS.461.3794O, 2017MNRAS.466.3103S, 2017MNRAS.469.4899M, 2018ApJ...854..120M, 2018PASJ...70S..28M, 2018MNRAS.474.1361P, 2019ApJ...875...63M, 2019MNRAS.483.2871D, 2019MNRAS.482.1352M, 2019MNRAS.484.1598B, 2019PASJ...71..107M}. The MORs are not the same for these cluster samples and mass proxies. However, one should expect the cosmological constraints from using these different MORs to be consistent. 

Currently, the state-of-the-art mass calibration of optically selected clusters is performed by the Dark Energy Survey (DES) with cluster catalogs from the redMaPPer cluster finder \citep{2014ApJ...785..104R, 2016ApJS..224....1R}, using the optical count of red galaxies, $\lambda$, as the mass proxy, and DES shear catalogs for the weak lensing mass calibration. The red galaxy count $\lambda$ (a.k.a. ``richness'') is computed as the sum of the membership probabilities of red galaxies within a cluster scale radius and brighter than some luminosity threshold, where the membership probabilities are assigned based on a model for the red-sequence as a function of the redshifts ($z$) and on a radial filter. In the most recent result from \cite{2019MNRAS.482.1352M} with DES Year 1 (Y1) data, the mass calibration was performed in the range of $0.2<z<0.65$ and $\lambda > 20$. For the last analysis with DES Y1 through Year 6 (Y6) data, we expect to have a cluster sample going to $z\sim 1$. 

The key assumption that redMaPPer uses to identify clusters is that each cluster has a well defined red sequence population. At low redshifts this is a powerful assumption and allows efficient cluster finding. It is unclear at what redshift all clusters gain a red sequence. The evolution of the red sequence is still a topic of considerable debate \citep{2009ApJ...706L.173B, 2014A&A...571A..99S, 2016MNRAS.458L..14F, 2019A&A...632A..80G, 2019ApJ...880L..14C}. Several competing or complementary processes are responsible for driving or ceasing the star formation in the member galaxies. The dominant processes are expected to differ across redshifts, stellar masses, halo masses and environment (e.g. \citealt{2016A&ARv..24...14O}). For this reason, it is still a challenge to model the red-sequence population, in particular, at high redshifts (e.g. \citealt{2014A&A...571A..99S, 2016ApJ...825..113D, 2017MNRAS.471.1671D, 2019ApJ...877...48C}).

The redMaPPer cluster catalogues contain clusters at $\lambda > 5$, but the DES does not use clusters at $5 < \lambda < 20$ for cosmology as this low richness sample is unreliable as many of the lowest-richness clusters are subject to strong projection effects in the line-of-sight. However, these low mass samples are very interesting for astrophysical studies (e.g. \citealt{2009ApJ...696..620C, 2018ARA&A..56..435W}). By definition these systems have few members and problems with Poisson statistical noise become important. However, even a modest shift towards lower $\lambda$ values could potentially have a significant impact on cosmology. This is a challenging regime but the potential impact makes it worth exploring alternative mass proxies that might be more robust against projection effects. Alternative optical mass proxies are possible \citep{2012A&A...548A..83A, 2017MNRAS.472.3246M, 2018MNRAS.474.1361P, 2019MNRAS.484.1598B, 2020MNRAS.493.4591P, 2020arXiv200512275S}. For example, one could incorporate the count of star-forming galaxies into the richness. This would be of particular interest at low masses and high redshifts.

In \cite{2018MNRAS.474.1361P} and \cite{2020MNRAS.493.4591P} we introduced and studied a physically motivated mass proxy named $\mu_{\star}$, which is based on the total stellar mass and therefore accounts for the red and blue members of the clusters. \citet{2012A&A...548A..83A} was the first to propose a stellar mass-based mass proxy for clusters, but since then such kind of proxy has mostly been studied in simulations \citep{2016MNRAS.456.4291A, 2017MNRAS.464.2270A, 2018AstL...44....8K, 2018MNRAS.478.2618F, 2020MNRAS.493..337B}. 

In particular, \cite{2020MNRAS.493..337B} showed that a stellar mass proxy similar to $\mu_{\star}$ has less intrinsic scatter with halo mass than a richness proxy and is less affected by projection effects. They used a set of simulations for this comparison, in which they identify halos and compute the intrinsic scatter in the virial mass at fixed proxy by: $i)$ using the true redshifts, i. e. no projection effects; $ii)$ simulating a spectroscopic survey with precise redshift measurements and $iii)$ simulating a photometric survey with redshift uncertainty of $\sigma_z/(1+z)=0.01$. In all these cases, they showed that the proxy based in the total stellar mass presented lower intrinsic scatter than the $\lambda$--like proxy (see their Figure 3).       

In \cite{2018MNRAS.474.1361P} we provided a first calibration of the mass--$\mu_{\star}$ relation at low $z$ using the SDSS Stripe 82 data. In this work, we use stacked weak lensing signal to measure the mean galaxy cluster mass of redMaPPer clusters identified in DES Y1 data using $\mu_{\star}$ as a mass proxy. For the first time, we calibrate the mass--$\mu_{\star}$--redshift relation of these clusters at moderate redshifts ($z \leq 0.7$). We also incorporate a variety of improvements to the weak lensing modelling and perform a blinded analysis.          

This paper is organised as follows. In \autoref{data}, we describe the cluster and the lensing shear catalogues. In \autoref{methodology}, we present the methodology for the measurement and modelling of the stacked cluster masses. We present the modelling and the derived mass-calibration in \autoref{sec:masscal} and \autoref{sec:results}, respectively. Finally, in \autoref{sec:disc}, we present our concluding remarks and we summarise our results in \autoref{sec:summ}. 

In this paper, the distances are expressed in physical coordinates, magnitudes are in the AB system (unless otherwise noted) and we denote logarithm base 10 as $\log$ and logarithm base $e$ as $\ln$. We assume a flat $\Lambda$CDM cosmology with a matter density $\Omega_{m}=0.3$ and a Hubble parameter $h = H_0/100\,\mathrm{km\,s^{-1} Mpc^{-1}} =1$. 

\section{The DES Y1 catalogues}
\label{data}

The Dark Energy Survey \citep{2005astro.ph.10346T, 2016MNRAS.460.1270D} is an optical imaging survey that observed 5,000 square degrees of the celestial southern hemisphere using the 4m Blanco Telescope and the Dark Energy Camera \citep[DECam;][]{2015AJ....150..150F} at the Cerro Tololo Inter-American Observatory (CTIO) in Chile. The main goal of the survey is to constrain the distribution of dark matter in the Universe, and the amount and properties of dark energy, including its equation of state. DES used the $grizY$ bands to obtain photometric redshifts and reaching limiting magnitudes of $i \sim 24$. Due to the large area, depth, and image quality of DES, we expect to have an optical identification of a large number of galaxy clusters and groups ($\sim 100,000$) up to a redshift $z\sim1$. 

The DES observations were carried out during roughly one semester per year, and the first full operating season took place from August 2013 to February 2014, DES Y1 \citep{2014SPIE.9149E..0VD, 2018ApJS..235...33D}. Before this, a small Science Verification (SV) survey was conducted from November 2012 to February 2013. The SV data covered 250 square degrees reaching almost the depth of the complete survey. 

During the DES Y1 observations, 1,839 square degrees of the southern sky were observed in three to four tilings in the $griz$ bands as well as 1,800 square degrees in the $Y$-band. The resulting imaging is shallower than the SV data but covers a significantly larger area. In the DES Y1 we have $\sim 1,500$ square degrees of the main survey, divided into two large non-contiguous areas. The reduction in the area is due to a series of survey masks. These masks are applied to avoid bright stars, satellite tracks, the Large Magellanic Cloud, among others. The two non-contiguous areas are the ``SPT'' area (1,321 square degrees), which overlaps the footprint of the 2,500 square degrees South Pole Telescope Sunyaev-Zel$^{\prime}$dovich Survey \citep{2011PASP..123..568C}, and the ``S82'' area (116 square degrees), which overlaps the Stripe-82 deep field of the Sloan Digital Sky Survey \citep[SDSS;][]{2014ApJ...794..120A}. In this study, we utilise data from the SPT region.  

The data from the first three seasons was the basis for the first DES public data release\footnote{ \url{https://www.darkenergysurvey.org/}} \citep{2018ApJS..239...18A}. The data processing for Y5 has already been completed and the final observing season, Y6, was finalised on January 9th, 2019. 

In the following, we briefly describe the catalogues used in this analysis and refer the reader to the corresponding papers for more details. The photometric redshift and weak lensing shape catalogues were used in the main DES cosmological analysis combining galaxy clustering and weak lensing \citep{2018PhRvD..98d3526A}.

\subsection{redMaPPer cluster catalogue}

In this work, we use the ``volume-limited'' catalogue of photometrically selected clusters identified in DES Y1 data by the redMaPPer cluster-finding algorithm v6.4.17 \citep{2014ApJ...785..104R, 2016ApJS..224....1R}. In this catalogue, a galaxy cluster is included in the sample only if all cluster member galaxies brighter than the luminosity threshold used to define cluster richness in redMaPPer are above $3\sigma$ limiting magnitude in $g$, $5\sigma$ in $r$ and $i$, $10\sigma$ in $z$ according to depth maps of the survey \citep{2018ApJS..235...33D}.

As previously mentioned, redMaPPer uses multiband colours to find overdensities of red-sequence galaxies around candidate central galaxies. In DES Y1 data, redMaPPer uses the four band magnitudes ($griz$) and their errors to spatially group the red-sequence galaxies at similar redshifts into cluster candidates. Starting from an initial set of spectroscopic seed galaxies, the algorithm iteratively fits a model for the local red sequence, and for each red galaxy, redMaPPer estimates its membership probability ($p_{\mathrm{mem}}$) following an iteratively self-trained matched-filter technique. At the end, for each identified cluster, redMaPPer returns an optical richness estimate $\lambda$ (the sum over the membership probabilities of all red galaxies within a pre-defined, richness-dependent projected radius $R_{\lambda} = (\lambda/100)^{0.2} h^{-1} \mathrm{Mpc}$), a photo-z estimate $z_{\lambda}$ (obtained by maximizing the probability that the observed colour-distribution of likely members matches the self-calibrated red-sequence model of redMaPPer), the positions (RA, Dec) and a vector with the probabilities of the five most likely central galaxies ($P_{\mathrm{cen}}$).

\begin{figure}
\includegraphics[width=\columnwidth]{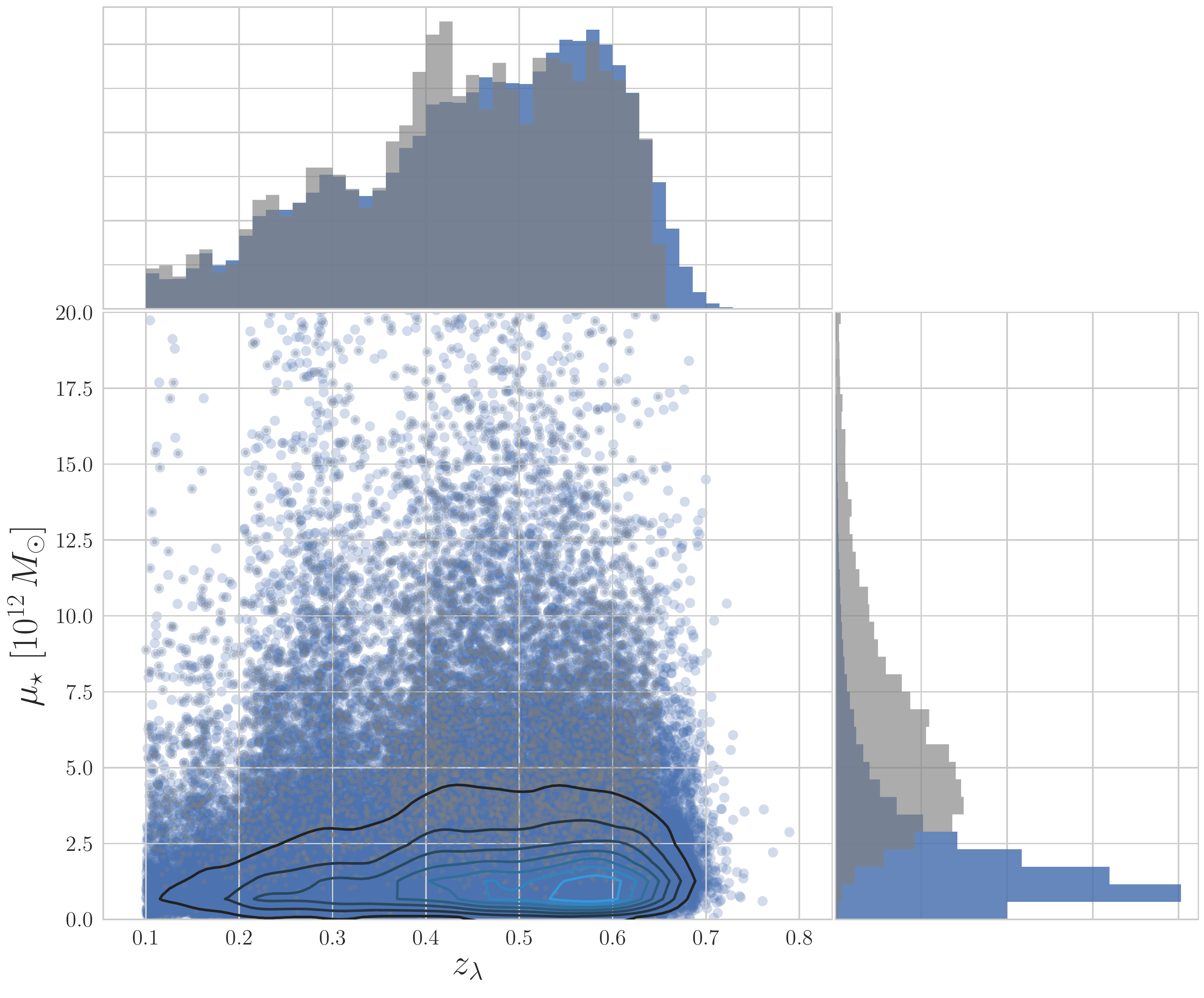}
\caption{Redshift--$\mu_{\star}$ distribution of the redMaPPer clusters in the volume limited DES Y1 cluster catalogue with $\lambda >5$ (\textit{blue} dots) overlapped with density contours to highlight the densest regions and the redshift-$\mu_{\star}$ distribution of the clusters ($\lambda >20$, $0.1\leq z<0.65$) used in this analysis (\textit{grey} dots). At the top and on the right are normed histograms of the projected quantities, $z_{\lambda}$ and $\mu_{\star}$, respectively, for the full catalogue (in \textit{blue}) and for the subsample used in this work (in \textit{grey}).}
\label{fig:muxzdist}
\end{figure}

This catalogue contains more than 76,000 clusters down to $\lambda>5$ and out to $z_{\lambda} \sim 0.8$, of which more than 6,000 are above $\lambda \geqslant 20$. For each cluster in this catalogue, we computed the value of the mass proxy $\mu_{\star}$, which will be described in the next section. In \autoref{fig:muxzdist} we show the cluster $\mu_{\star}$ and redshift distributions for the volume-limited catalog ($\lambda>5$) in \textit{blue}. Because the spectroscopic training sample goes only to $z\sim 0.65$, the catalogue should be robust just within this range. To avoid complications with selection functions and unreliable detections due to projection effects, in this work we just use the sample with $\lambda>20$, which is the sample used in the main cosmology analysis of DES. Thus, in \textit{grey} we show the cluster sample we use: $0.1<z<0.65$ and $\mu_{\star}<5.5\times 10^{13}\, \mathrm{M_{\odot}}$ with a total of 6,124 galaxy clusters.        

\begin{figure}
\includegraphics[width=\columnwidth]{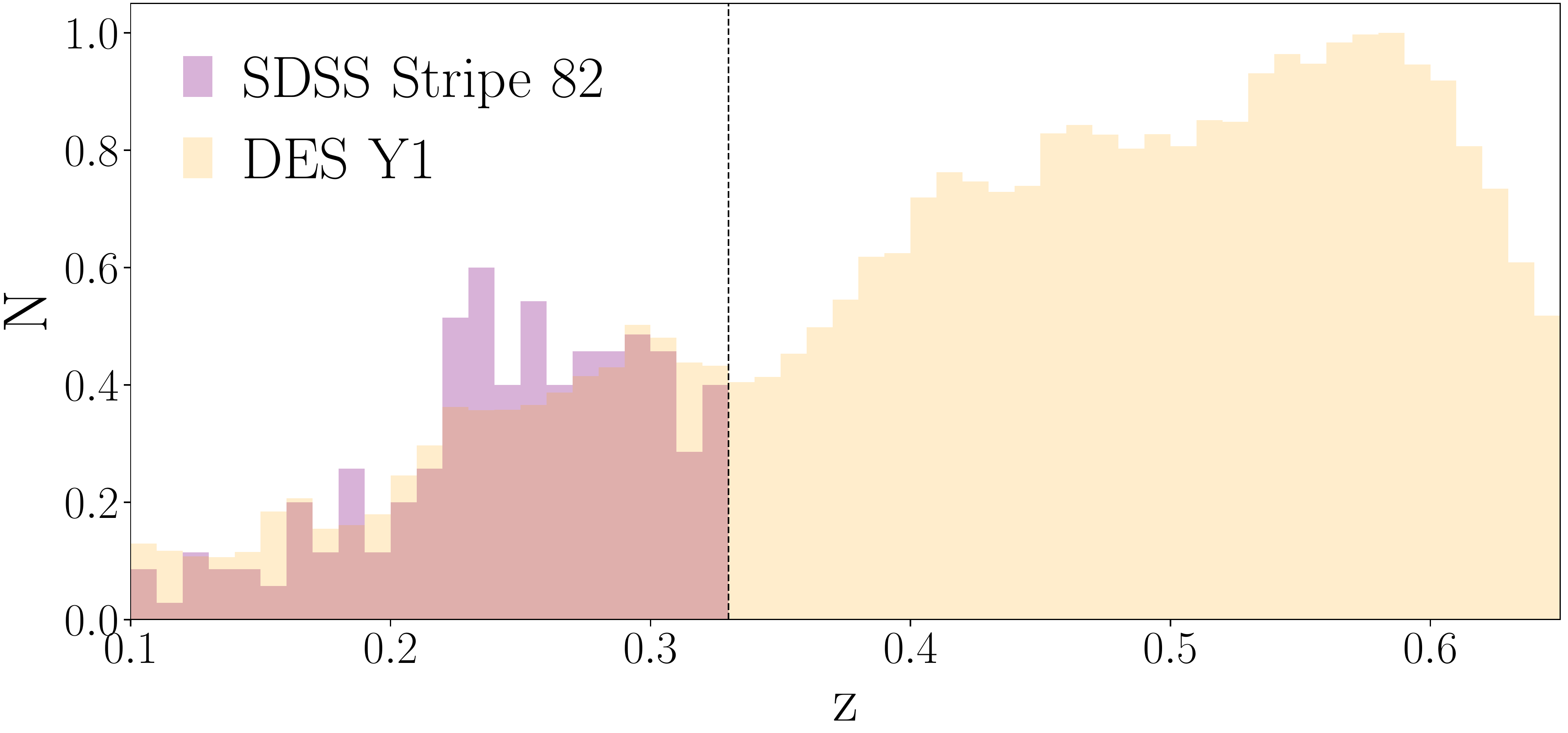}
\caption{Normalised redshift distribution: 230 clusters in the SDSS Stripe 82 sample (\textit{purple}) and 6,124 clusters in DES Y1 sample (\textit{orange}). In both histograms, the catalogue used corresponds to the $\lambda > 20$ samples. The vertical dotted line shows the $z$ limit ($z=0.33$) in which the previous calibration of $\mu_{\star}$ has been performed.}
\label{fig:zdist}
\end{figure}

In \cite{2018MNRAS.474.1361P} we performed the first mass-calibration of $\mu_{\star}$ for the SDSS Stripe 82 redMaPPer catalogue of $\lambda >20$ and $0.1<z<0.33$ with a total of 230 clusters and using shear data from the CFHT Stripe 82 Survey \citep[CS82;][]{2014RMxAC..44..202M}. In \autoref{fig:zdist}, we show the comparison of the normalised redshift distribution of the SDSS Stripe 82 sample (\textit{purple}) with the DES Y1 sample (\textit{orange}) of $\lambda >20$ and $0.1<z<0.65$. It illustrates the increase towards higher redshifts and the statistical gain compared to the previous work, which allows to study the redshift evolution of our MOR and reduce the statistical errors in our mass measurements.

\subsection{Photometric redshift catalogue}

To estimate physical quantities such as $\mu_{\star}$ and the mass from the lensing signal we need to know the redshift of the member galaxies and the source galaxies being lensed, respectively. We also need the information about the individual $P(z)$ of source galaxies for computing the boost-factors profiles.

We use the photometric redshift catalogue\footnote{\url{https://des.ncsa.illinois.edu/releases/y1a1/key-catalogs/key-photoz}} of DES Y1 produced and validated by \cite{2018MNRAS.478..592H} using the template-based BPZ\footnote{\url{http://www.stsci.edu/~dcoe/BPZ/}} algorithm \citep{2000ApJ...536..571B, 2006AJ....132..926C}. \cite{2018MNRAS.478..592H} found that these photo-z estimates were biased and needed an overall multiplicative systematic correction in the recovered weak lensing profiles. Following \cite{2019MNRAS.482.1352M} we determine this correction and present it in \autoref{methodology}. We also use two separate BPZ catalogues: one generated from the single epoch \textsc{metacalibration}-measured photometry for selecting and weighting sources, and one from a multi-epoch, multi-object fitting (MOF) photometry for determining the resulting source redshift distributions.

\subsection{Assigning \texorpdfstring{$\mu_\star$}{m} for redMaPPer clusters}

The $\mu_{\star}$ mass proxy relies on the clear physical meaning of the total stellar mass of a cluster and, in particular, \cite{2020MNRAS.493.4591P} showed that the scatter in the $\mu_{\star}$ to X-ray temperature relation is comparable to other mass proxies (in particular, to the redMaPPer $\lambda$), for an X-ray sample. 

In \cite{2018MNRAS.474.1361P} and \cite{2020MNRAS.493.4591P} we describe in detail how the mass proxy $\mu_{\star}$ is computed. The assignment of $\mu_{\star}$ is the last step in the modular pipeline \texttt{vt-clustertools} \href{https://github.com/SSantosLab/vt-clustertools}{\faGithubAlt} that we are developing and which consists of: $i)$ cluster finding (optional, since any cluster catalogue can be given as input); $ii)$ probabilistic membership assignment; $iii)$ total stellar-mass measurements and $\mu_{\star}$ estimation. 

We use DES Y1 redMaPPer catalogue as input and performed the membership assignment and stellar mass measurements. First, we compute the membership probability $P_{\mathrm{mem}}$ for each cluster galaxy as  
\begin{equation}
P_{\mathrm{mem}} = P_\mathrm{z} P_\mathrm{r},
\end{equation}
where the two components represent the probability of the galaxy to be a member given its redshift ($P_\mathrm{z}$) and its distance from the cluster centre ($P_\mathrm{r}$). In practice, $P_z$ is the integrated photometric redshift probability distribution of each galaxy within a $\Delta z=0.1$ window around the cluster redshift. $P_r$ is computed assuming a projected Navarro-Frenk-White profile from \cite{1999astro.ph..8213O}, where $r_{200}$ is defined as the radius at which the mass density of the cluster is 200 times greater than the critical density of the universe $\rho_{\mathrm{crit}}=3H^2(z)/8\pi G$, where $H(z)$ is the respective Hubble parameter. For the concentration parameter we assume a fixed value of $c=3$.   

After computing the membership probabilities for each galaxy $i$ within 3 Mpc of each cluster $j$, we compute their stellar masses assuming that every member galaxy is at the redshift of its host, $M_{\star,i}(z_j)$. Since the cluster redshifts have smaller uncertainties than individual galaxies' redshifts, this minimizes the uncertainties on $M_{\star,i}$ measurements. The stellar masses are computed using the Bayesian Model Averaging (BMA; \citealt{hoeting}) method, as described in \cite{2020MNRAS.493.4591P}. Once the stellar masses are computed, we define the mass proxy $\mu_{\star}$ as the sum of the individual galaxy stellar masses weighted by their membership probability
\begin{equation}
\mu_{\star} = \sum_{i} P_{\mathrm{mem},i} M_{\star,i} \,.
\end{equation}

The membership assignment and $\mu_{\star}$ computation were performed in the full DES Y1 volume-limited catalogue with $\lambda>5$, but through this work we only use the clusters with $\lambda>20$ to make sure our analysis is done in the same regime as the current $\lambda$-sample to facilitate comparisons between the two mass proxies. 

\subsection{Weak lensing shear catalogue}

We use the shape measurements from the \textsc{metacalibration} \citep{2017ApJ...841...24S, 2017arXiv170202600H} shape catalogue\footnote{\url{https://des.ncsa.illinois.edu/releases/y1a1/key-catalogs/key-shape}} of DES Y1 presented in \cite{2018MNRAS.481.1149Z}. The \texttt{Metacalibration} \href{https://github.com/esheldon/ngmix/wiki/Metacalibration}{\faGithubAlt} code utilises images taken in $riz$ bands to measure the ellipticities of the galaxies. The algorithm works by distorting the image with a small known shear and calculating the response of a shear estimator to that applied shear. In this method, there is no need for prior information about galaxy properties or a calibration from simulations. The fiducial shear estimates are obtained from a single Gaussian fit by using the \texttt{ngmix} \href{https://github.com/esheldon/ngmix}{\faGithubAlt} model-fitting algorithm \citep{2015ascl.soft08008S}. The produced DES Y1 \textsc{metacalibration} catalog has an effective source density of 6.28 arcmin$^{-2}$.

The main systematic effect in this shape estimation is a multiplicative bias, i.e. an over- or underestimation of the gravitational shear inferred from the mean tangential ellipticity of the lensed galaxies. To characterize and correct for this bias, \texttt{Metacalibration} uses the galaxy images themselves to ``de-bias'' the shear estimates.

The \textsc{metacalibration} shear catalogue and the associated calibration of the source redshift distributions \citep{2018MNRAS.478..592H} was extensively tested and validated by \cite{2018MNRAS.481.1149Z} and \cite{2018PhRvD..98d2005P}, making this lensing catalogue well tested for different applications. 

\section{The stacked cluster profiles}
\label{methodology}

We measure the mass of the DES Y1 redMaPPer clusters from their stacked weak lensing signal using the \textsc{metacalibration} shear catalogue and BPZ photo-z's. For the stacking of the lenses, we define bins of redshift and $\mu_{\star}$. The redshift bins are defined as $z_{\mathrm{low}}$ ($0.1 \leqslant z < 0.33$), $z_{\mathrm{mid}}$ ($0.33 \leqslant z < 0.5$) and $z_{\mathrm{high}}$ ($0.5 \leqslant z < 0.65$). To remove the few outlier cases in which the stellar-mass fitting code generated nonphysical values for one or more of the member galaxies, we apply an upper limit cut in the $\mu_{\star}$ range such that the clusters in these three redshift bins lie within the range $\mu_{\star} < 5.5 \times 10^{13} M_{\odot}$. In each redshift bin, we separate the samples into four $\mu_{\star}$ bins, in such a way that we have a similar number of clusters in each bin. In \autoref{tab_cluster_binnig}, we summarise the binning scheme for our stacking measurements.  

\begin{table}
	\centering
	\caption{Binning scheme and properties of the DES Y1 redMaPPer cluster sample. We split the clusters into three redshift bins and choose the $\mu_{\star}$ bins to have a similar number of clusters in each of the four bins. Here $\mu_{\star}$ is in units of $10^{12} \, \mathrm{M_{\odot}}$.}
	\label{tab_cluster_binnig}
	\begin{tabular}{cclcc} %
		\hline
		$z$ range & Mean $z$ & $\mu_{\star}$ range & Mean $\mu_{\star}$ & No. of clusters\\
		\hline
		\multirow{4}{*}{[0.1, 0.33)} & 0.232 & $\left[0, 3.3\right)$ & 2.38 & 318\\
		 & 0.233 & $\left[3.3, 5.0\right)$ & 4.10 & 317\\
		 & 0.243 & $\left[5.0, 7.5\right)$ & 6.15 & 313\\
		 & 0.259 & $\left[7.5, 60\right)$ & 12.6 & 402\\
		\hline
		\multirow{4}{*}{[0.33, 0.5)} & 0.424 & $\left[0, 4.04\right)$ & 3.06 & 571\\
		 & 0.420 & $\left[4.04, 5.65\right)$ & 4.83 & 567\\
		 & 0.420 & $\left[5.65, 8.05\right)$ & 6.73 & 567\\
		 & 0.427 & $\left[8.05, 60\right)$ & 12.98 & 744\\
		\hline
        \multirow{4}{*}{[0.5, 0.65)} & 0.572 & $\left[0, 3.88\right)$ & 2.96 & 554\\
		 & 0.574 & $\left[3.88, 5.42\right)$ & 4.67 & 555\\
		 & 0.573 & $\left[5.42, 7.68\right)$ & 6.46 & 556\\
		 & 0.570 & $\left[7.68, 60\right)$ & 11.71 & 660\\
		\hline

	\end{tabular}    
\end{table}

\subsection{Projected surface mass density profiles}
\label{projected-ds}

In the weak lensing regime, a non-linear combination of the gravitational shear $\gamma$ and convergence $\kappa$ defines an estimator for the ``reduced shear'' \citep{2001PhR...340..291B}
\begin{equation}
\label{eq:redu_shear}
 \mathbf{g} \equiv \frac{\gamma}{1-\kappa}.    
\end{equation}
 In practice, we assume $\langle \mathbf{g} \rangle \approx \langle \mathbf{\gamma} \rangle \approx \langle \mathsf{R} \rangle^{-1} \langle \mathbf{e} \rangle$. 
 Here $\mathsf{R}$ is a joint response matrix computed as $\mathsf{R} \approx \mathsf{R}_{\gamma} + \mathsf{R}_{\mathrm{sel}}$, where the terms on the right are the responses of the ellipticity measurement and the selection effects to the gravitational shear, respectively (see \citealt{2017ApJ...841...24S} and \citealt{2019MNRAS.482.1352M} for details).  
    
The gravitational field from a foreground mass distribution induces correlations in the shapes of source galaxies, such that, on average, galaxies images are stretched and aligned tangentially to the centre of mass. \cite{1991ApJ...370....1M} found that, for any distribution of projected mass, it is possible to show that the azimuthally averaged tangential shear $\gamma^{\mathsf{T}}$ at a projected radius $R$ from the centre of the mass distribution is given by 
\begin{equation}
\gamma^{\mathsf{T}} (R) = \frac{\Delta\Sigma}{\Sigma_{\mathrm{crit}}} \equiv \frac{\overline{\Sigma}(<R) - \overline{\Sigma}(R)}{\Sigma_{\mathrm{crit}}},
\label{deltasigmat}
\end{equation}
where $\Sigma(R)$ is the projected surface mass density at radius $R$, $\overline{\Sigma}(<R)$ is the mean value of $\Sigma$ within a disc of radius $R$ given by
\begin{equation}
  \label{eq:barSigma}
  \overline{\Sigma}(<R) = \frac{2}{R^2}\int_0^R{\rm d}R'\ R' \overline{\Sigma}(R'),
\end{equation}
and $ \overline{\Sigma}(R)$ is the azimuthally averaged $\Sigma (R)$ within a ring of radius $R$ computed as
\begin{equation}
  \label{eq:Sigma}
  \overline{\Sigma}(R) = \int_{-\infty}^{+\infty}{\rm d}\chi\ \Delta\rho\left(\sqrt{R^2+\chi^2}\right)\,,
\end{equation}
where $\chi$ is the separation along the line of sight and $\Delta \rho$ is an average excess of a given three-dimensional matter density. Finally, $\Sigma_{\mathrm{crit}}$ is the critical surface mass density expressed in physical coordinates as 
\begin{equation}
\Sigma_{\mathrm{crit}} = \frac{c^2 D_s}{4 \pi G D_l D_{ls}},
\label{sigma_crit}
\end{equation}
where $D_l$ and $D_s$ are angular diameter distances from the observer to the lens and to the source, respectively, and $D_{ls}$ is the angular diameter distance between lens and source.

To perform precise measurements of the surface density contrast $\Delta\Sigma$, we need to estimate the redshifts of the lens (i.e. galaxy clusters) and the source galaxies robustly. We use the photometric redshift estimates from the redMaPPer algorithm as the lens redshifts. Due to a negligible statistical uncertainty on these estimates ($\Delta z_\mathrm{l} \approx 0.01$, \citealt{2016ApJS..224....1R}), compared to other sources of error in the lensing measurement, we can treat these redshifts as exact. The redshift of source galaxies are also photometric, and are described by a probability distribution $p_{\mathrm{phot}}(z_\mathrm{s})$ for each source galaxy. Therefore, we estimate an effective critical surface density for each lens-source pair
\begin{equation}
\langle \Sigma_{\mathrm{crit}}^{-1} \rangle_{j,i} = \int dz_{\mathrm{s},i} p_{\mathrm{phot}}(z_{\mathrm{s},i}) \Sigma_{\mathrm{crit}}^{-1} (z_{\mathrm{s},i}, z_{\mathrm{l},j}),   
\label{pphotz}
\end{equation}
that averages over the $p_{\mathrm{phot}}(z_{\mathrm{s},i})$ of source $i$, evaluated for lens $j$. For computational reasons, we do not use the full integral over $p_{\mathrm{phot}}(z_{\mathrm{s},i})$, but rather replace \autoref{pphotz} by $\Sigma_{\mathrm{crit}}^{-1}$ evaluated at a random sample of the $p_{\mathrm{phot}}(z_{\mathrm{s}})$. This kind of approximation is justified in \cite{2018AJ....156...35M}, for example. 

From \autoref{deltasigmat}, we can compute $\Delta\Sigma$ over several lenses with similar physical properties (e.g. redshift, stellar mass) to increase the signal-to-noise and average over the effect of substructures, uncorrelated structures in the line of sight, shape noise and variations in the shape of individual halos. However, in practice, using the shear and selection responses ($\mathsf{R}_{\gamma}$ and $\mathsf{R}_{\rm sel}$, respectively) provided in \textsc{metacalibration}'s catalogue we define a minimum variance estimator for the weak lensing signal as    
\begin{equation}
	\label{eq:updated_deltasigma_estimate}
	\widetilde{\Delta\Sigma} \; \equiv\; \frac{\sum\limits_{j,i} \omega_{i,j}\;  e_{{\rm T};\,i,j} }{\sum\limits_{j,i} \omega_{i,j}\, \Sigma_{{\rm crit;}i,j}^{'-1}\, \mathsf{R}^{\rm T}_{\gamma,i}  + \left(\sum\limits_{j,i} \omega_{i,j}\, \Sigma_{{\rm crit;}i,j}^{'-1}\right)\langle \mathsf{R}^{\rm T}_\mathrm{sel} \rangle } \; ,
\end{equation}
where the summation goes over all source-lens pair in a given radial bin and $e_{\mathrm{T};i,j}$ is the tangential component of source $i$ relative to the lens $j$. The quantities $\mathsf{R}^{\rm T}_{\gamma,i}$ and $\langle \mathsf{R}^{\rm T}_\mathrm{sel} \rangle$ are proportional to the trace of the shear and selection response matrices, respectively, and their detailed definitions can be found in \cite{2018MNRAS.481.1149Z} and \cite{2019MNRAS.482.1352M}, but it is important to note that these selection responses were defined by the photometric redshift estimates derived from the sheared METACALIBRATION photometry. 

To speed up the computation of \autoref{eq:updated_deltasigma_estimate}, we use two simplifications: $i)$ replace the expectation value of the normalisation $\Sigma_{{\rm crit}}^{'-1}$ by a Monte Carlo estimate     
\begin{equation}
	\label{eq:sigma_crit_mc_estimate}
	\Sigma_{{\rm crit;}i,j}^{'-1}=\Sigma_{\rm crit}(z_{{\rm l}_j},z_{{\rm s}_i}^{\rm MC}) \,,
\end{equation}
where $z_{s_i}^{\rm MC}$ is a random sample from the $p_\mathrm{phot}(z_s)$ distribution estimated with BPZ using MOF photometry; $ii)$ choose the weights as 
\begin{equation}
	\label{eq:updated_weights}
	\omega_{i,j} \equiv \Sigma_{{\rm crit}}^{-1}\left(z_{{\rm l}_j}, \langle z_{{\rm s}_i}^{{\rm MCAL}}\rangle\right)\; \mathrm{if}\; \langle z_{{\rm s}_i}^{{\rm MCAL}}\rangle > z_{{\rm l}_j} + \Delta z\,,
\end{equation}
where $\langle z_{\rm s_i}^{{\rm MCAL}}\rangle$ is the mean redshift of the source galaxy estimated from \textsc{metacalibration} photometry. We use separation of $\Delta_z=0.1$ from the lens-redshift for source selection. \cite{2019MNRAS.482.1352M} found that including the source weights provided by \textsc{metacalibration} does not introduce a significant improvement in the signal-to-noise of the measurement. They also argue that the use of two different photometric estimators is necessary because when calculating the selection response, the internal photometry of the \textsc{metacalibration} must be used for all selections and weightings of sources. 

In addtion to that, \cite{2018MNRAS.478..592H} found that photo-z estimates from \textsc{metacalibration} have a greater scatter than the ones estimated with MOF photometry. Therefore, we follow the approach of \cite{2019MNRAS.482.1352M} expressed in \autoref{eq:updated_deltasigma_estimate}, where we use the \textsc{metacalibration} photo-z estimates for selecting and weighting the source-lens pairs and we use the MOF-based photo-z estimates for computing the normalisation of the shear signal to find $\Delta\Sigma$.    

\begin{figure*}
\includegraphics[width=\linewidth]{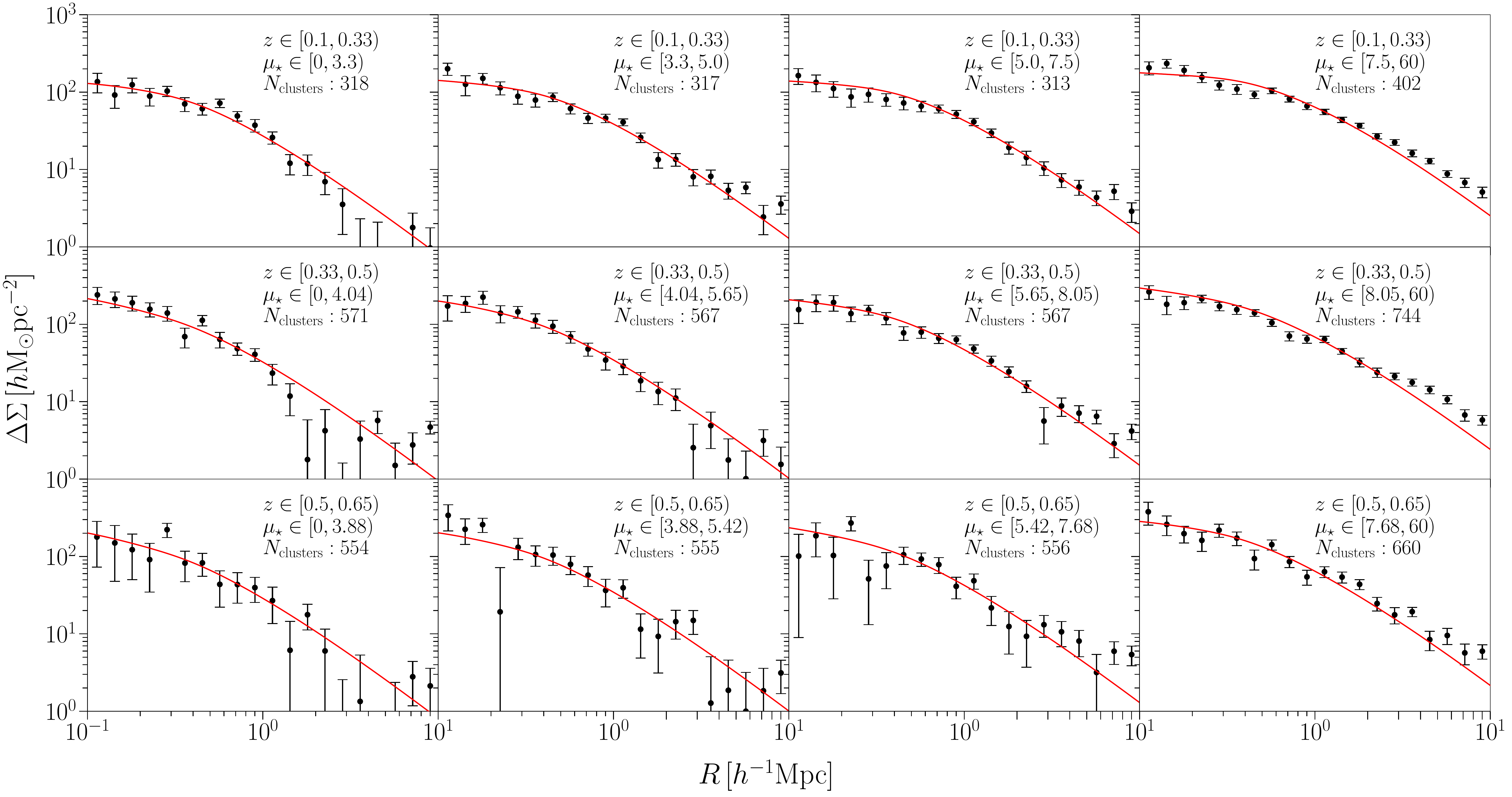}
\caption{The mean $\Delta\Sigma$ for cluster subsets separated in $z_l$ (increasing from \textit{top} to \textit{bottom}) and $\mu_{\star}$ (increasing from \textit{left} to \textit{right}) with errors from jackknife resampling (see \autoref{sec:covs}). The best-fit model (\textit{red curve}) includes the corrections in \autoref{eq:full_delta_sigma_model}, see \autoref{methodology} for details. In the legend, $\mu_{\star}$ is in units of $10^{12}\, \mathrm{M_{\odot}}$.}
\label{fig:des_ds}
\end{figure*}

To estimate the weak lensing signal $\Delta\Sigma$ from \autoref{eq:updated_deltasigma_estimate}, we use a modified version of the \texttt{xshear} \href{https://github.com/esheldon/xshear}{\faGithubAlt} code implemented in the \texttt{xpipe} \href{https://github.com/vargatn/xpipe}{\faGithubAlt} Python package. The clusters are grouped into three bins in redshift: $z\in[0.1;0.33)$, $[0.33;0.5)$, and $[0.5;0.65)$, as well as four bins in $\mu_{\star}$ as described in \autoref{tab_cluster_binnig}. We measure the $\Delta\Sigma$ profiles in 20 logarithmic radial bins in the range $(0.1-10)\, h^{-1}$ Mpc. The measured $\Delta\Sigma$ profiles are shown in \autoref{fig:des_ds}. We computed the cross-component of the lensing signal ($\Delta\Sigma_{\times}$) and found no evidence of spurious correlations in the weak-lensing signals, i.e. the measured $\Delta\Sigma_{\times}$ are consistent with zero. \cite{2019MNRAS.482.1352M} described a series of tests and validation for systematics of the source catalogue such as shear and photometric redshift bias and cluster members contamination. Since we rely on the same catalogue, the treatment of this systematics could be applied to our work and is described in detail in the next sections.      

\subsection{Covariance matrices for $\Delta\Sigma$} 
\label{sec:covs}

In \cite{2018MNRAS.474.1361P}, the measurements were shape-noise dominated such that the covariance between adjacent radial bins was not noticeable and the $\Delta\Sigma$ measurements in each bin were treated as independent. However, for the DES Y1 sample, this assumption does not hold anymore. Besides the shape noise, the uncertainty in the $\Delta\Sigma$ measurements have contributions from the uncertainty in the photometric redshift estimations, and the intrinsic variations of cluster profiles. Furthermore, in a stacked cluster lensing analysis in a given survey area, source galaxies are paired with multiple clusters, possibly generating covariance between different radial bins as well as different cluster bins in $\mu_{\star}$ and redshift. Therefore, the cluster $\Delta\Sigma$  measurements are not fully independent, and we need to estimate the covariance matrix $\mathsf{C}_{\widetilde{\Delta\Sigma}}$ that will have significant off-diagonal terms, in particular, on large scales.  

Following \cite{2019MNRAS.482.1352M}, to estimate $\mathsf{C}_{\widetilde{\Delta\Sigma}}$ we use a spatial jackknife (JK) scheme designed to account for the covariance of the measurements. We use a JK resampling with $K=100$ simply-connected spatial regions $\mathcal{R}_k$ selected by running a \texttt{k-means} \href{https://github.com/esheldon/kmeans_radec}{\faGithubAlt} algorithm on the sphere. The JK covariance is defined as in \cite{1982jbor.book.....E} by
\begin{equation}
\label{eq:jackknife_cov}
C_{\widetilde{\Delta\Sigma}} = \frac{K - 1}{K} \sum\limits_{k}^{K}\left(\widetilde{\Delta\Sigma}_{(k)} - \widetilde{\Delta\Sigma}_{(\cdot)}\right)^T \cdot \left(\widetilde{\Delta\Sigma}_{(k)} - \widetilde{\Delta\Sigma}_{(\cdot)}\right)\,,
\end{equation}
where $\widetilde{\Delta\Sigma}_{(\cdot)} = \frac{1}{K}\sum_k \widetilde{\Delta\Sigma}_{(k)}$, and $\widetilde{\Delta\Sigma}_{(k)}$ is the lensing signal estimated through \autoref{eq:updated_deltasigma_estimate}, using all lenses except those in the region $\mathcal{R}_k$.
\begin{figure}
\includegraphics[width=\linewidth]{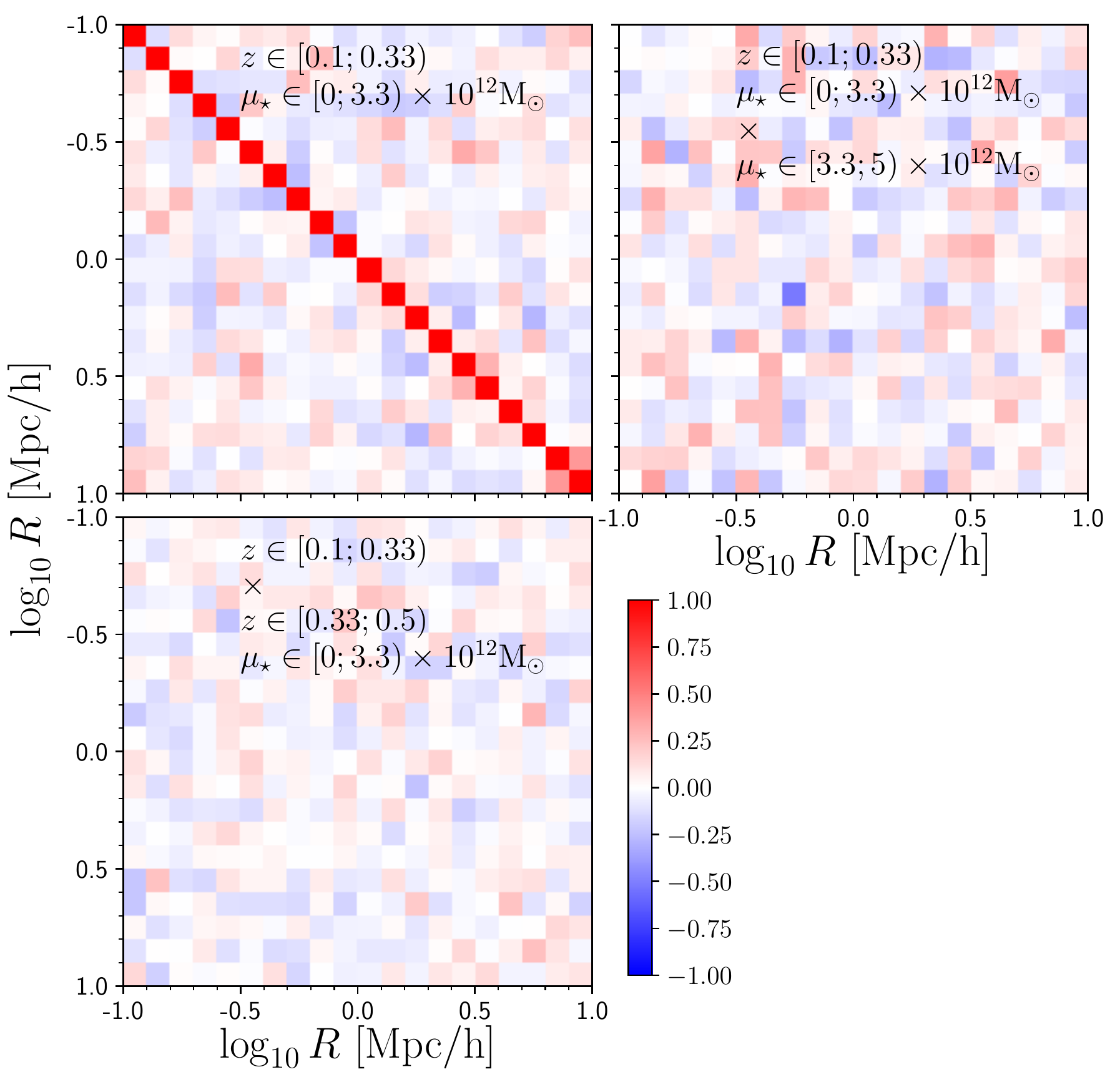}
\caption{Correlation matrix of $\widetilde{\Delta\Sigma}$ of a single profile with $z\in[0.1;0.33)$ and $\mu_{\star}\in[0;3.3)\times 10^{12} \mathrm{M_{\odot}}$, estimated from 100 jackknife regions (\textit{upper left} panel). The off-diagonal blocks show the correlation matrix between the reference profile and the neighbouring bin, $\mu_{\star}\in[3.3;5.0)\times 10^{12} \mathrm{M_{\odot}}$ (\textit{upper right} panel), and the neighbouring redshift bin $z\in[0.33, 0.5)$ (\textit{lower left} panel).}
\label{fig:des_corr_ds}
\end{figure}
In \autoref{fig:des_corr_ds}, we present an example of the estimated JK correlation matrix for the bin $z\in[0.1;0.33)$ and $\mu_{\star}\in[0;3.3)\times10^{12}\mathrm{M_{\odot}}$. We can see that on smaller scales the diagonal is dominant, but off-diagonal terms are present for the largest scales, as expected. We check the cross-correlations between the profiles measured in different $z$ and $\mu_{\star}$ subsets, and find them to be small (cf. upper right and lower left panel of \autoref{fig:des_corr_ds}). Therefore, we will assume no cross-correlation between different cluster subsets in our likelihood for fitting $\Delta\Sigma$.

\subsection{Projected surface mass density model}
\label{sec:surf_dens}

To infer the average masses from the weak lensing signal around each lens we use a two-components model given by a perfectly centred dark matter halo profile and an offsetted profile where the assumed centre does not correspond to the dynamical centre of the dark matter halo (``\textit{miscentring term}''), such that our $\Delta\Sigma$ model is given by 
\begin{equation}
\label{eq:two_ds}
\Delta\Sigma_{\rm model} = p_{\rm cc} \Delta\Sigma_{\rm cen} + (1 - p_{\rm cc}) \Delta\Sigma_{\rm misc},     
\end{equation}
where $p_{\rm cc}$ is the fraction of correctly centred clusters. For the centred profile, we could also consider the contribution of neighbouring halos through the ``2-halo'' term for the outer regions of the halo. However, for computational reasons, we choose to apply a radial cut ($R<2.5$ Mpc) while performing the profile-fitting of $\Delta\Sigma$ to minimise the effects of the 2-halo term. We have tested in simulations that this approach has negligible effects in the amplitude of the recovered $\Delta\Sigma$ (i.e. changes of 1--3 per cent).  

We model the centred term as a Navarro-Frenk-White (NFW; \citealt{1996ApJ...462..563N}) three-dimensional density profile given by
\begin{equation}
 \rho (r) = \frac{\delta_c \rho_{\mathrm{crit}}}{\frac{r}{r_s} \left( 1 + \frac{r}{r_s}\right)^2}, 
\end{equation}
where $r_s$ is the cluster scale radius, $\delta_{c}$ is the characteristic halo overdensity and $\rho_{\mathrm{crit}}$ is the critical density of the Universe at the lens redshift.

In this paper, we use the mass $M_{200}$ contained within a radius $r_{200}$ where the mean mass density is 200 times the critical density of the Universe. The scale radius is given by $r_s = r_{200}/c_{200}$, where $c$ is the concentration parameter. In our fitting procedure, we fix $c$ by assuming the semi-analytic concentration model of \cite{Diemer_2019} available in the \texttt{Colossus}  \href{https://bitbucket.org/bdiemer/colossus}{\faBitbucket} \citep{2015ascl.soft01016D} Python package. 

\cite{1996A&A...313..697B} and \cite{1999astro.ph..8213O} provide an analytical expression for the projected NFW profile, $\Delta\Sigma_{\mathrm{NFW}}$, and we use the Python code \texttt{NFW} \href{https://github.com/joergdietrich/NFW}{\faGithubAlt} \citep{jorg_dietrich_2016_50664} that implements these results for our profile-fitting procedure. Thus, the centred term in \autoref{eq:two_ds} is given by this $\Delta\Sigma_{\mathrm{NFW}}$. In the next section, we describe our model for the miscentring term, i.e. the $\Delta\Sigma_{\mathrm{misc}}$ for NFW density profiles.              

\subsubsection{Miscentring modelling}
\label{sec:misc_corr}

Miscentring can be caused by a simple failure in the centre assignment by the cluster finder algorithm. Also, many cluster finders assume as centre the position of the \textit{brightest cluster galaxy} (BCG). \cite{2012MNRAS.426.2944Z} show that some BCGs present an offset from the centre of their host dark matter halo. The redMaPPer code does not assume, necessarily, the position of the BCG as the cluster centre. Instead, redMaPPer uses a probabilistic approach to identify the top 5 most likely central candidates. Thus, the cluster position is given by the highest likelihood central galaxy. However, \cite{2016ApJS..224....1R} found that $\sim 80-85$ per cent of the redMaPPer central galaxies are BCGs and then subject to miscentring. In fact, \cite{2019MNRAS.tmp.1291Z} using high quality X-ray data found that $75 \pm 8$ per cent of redMaPPer clusters are well centred.             
The miscentring affects the observed shear profile \citep{2006MNRAS.373.1159Y,2007arXiv0709.1159J,2014MNRAS.439.3755F} and should be corrected. Therefore, we should estimate the miscentred differential mass density profiles as
\begin{equation}
\label{dsmisc_expression}
\Delta\Sigma_{\mathrm{misc}} (R) = \overline{\Sigma}_{ \mathrm{misc} }(<R) - \overline{\Sigma}_{\mathrm{misc}}(R)  
\end{equation}{}

We follow the modelling scheme presented in \cite{2007arXiv0709.1159J, 2012ApJ...757....2G, 2015MNRAS.447.1304F, 2017MNRAS.466.3103S,2018MNRAS.474.1361P} to compute the terms in \autoref{dsmisc_expression}. For a 2-dimensional offset in the lens plane $R_s$, the azimuthal average of the profile is 
\begin{equation}
\label{eq:sigma_mis}
\overline{\Sigma}_{\mathrm{misc}}(R) = \int_{0}^{\infty} dR_s P(R_s) \Sigma(R|R_s),  
\end{equation}
where
\begin{equation}
\Sigma(R|R_s) = \frac{1}{2\pi} \int_0^{2\pi} d\theta \Sigma\left(\sqrt{R^2+R_s^2+2RR_s\cos\theta}\right). 
\end{equation}
That is, the angular integral of the profile $\Sigma (R)$ is shifted by $R_s$ from the centre. The probability distribution of $R_s$ is given by
\begin{equation}
P(R_s) = \frac{R_s}{\sigma_{\mathrm{off}}^2} \exp\left(-\frac{1}{2} \frac{R_s^2}{\sigma_{\mathrm{off}}^2}\right),
\end{equation}
which is an \textit{ansatz} assuming the mismatching between the centre and $R_s$ follows a Rayleigh distribution. The mean surface density inside the radius $R$ is
\begin{equation}
\overline{\Sigma}_{ \mathrm{misc}} (<R) = \frac{2}{R^2}  \int_{0}^{R} dR^{\prime} R^{\prime} \overline{\Sigma}_{\mathrm{misc}}(R^{\prime}).
\end{equation}

We use the Python code \texttt{cluster-lensing} \href{https://github.com/jesford/cluster-lensing}{\faGithubAlt} \citep{2016AJ....152..228F, clusterlensing} that implements the equations 15--19 to compute the miscentring term $\Delta\Sigma_{\rm misc}(R)$ for NFW profiles. For this miscentring profile we only have one free parameter, the width of the offset distribution $\sigma_{ \mathrm{off}}$. Together with the parameter $p_{\mathrm{cc}}$ in \autoref{eq:two_ds}, i. e. the  fraction of correctly centred clusters, we would have two free parameters in our miscentring modelling.     

We decided to fix $\sigma_{\mathrm{off}}$ with the typical value derived by \cite{2019MNRAS.tmp.1291Z} for the DES Y1 redMaPPer clusters. In that work, they use a Gaussian instead of a Rayleigh distribution to model the distribution of offsets. However, our miscentring parameter $\sigma_{\mathrm{off}}$ is connected to their parameter $\tau$ as
\begin{equation}
\sigma_{\mathrm{off}} = \tau \times R_{\lambda}.    
\end{equation}{}
\cite{2019MNRAS.tmp.1291Z} found that $\tau = 0.17$ and in our sample, the average value of $R_{\lambda}$ is 0.78 $h^{-1} \mathrm{Mpc}$. Therefore, we fix $\sigma_{\mathrm{off}} = 0.133 h^{-1} \mathrm{Mpc}$ and keep $p_{\mathrm{cc}}$ as a free parameter when performing the profile-fitting.     

\subsubsection{Boost-factor model}
\label{sec:mod_boost}

The lensing signal can be diluted due to errors in the photometric redshift estimates that can cause some of our background sources to be either in the foreground ($z_s < z_l$) or to be physically associated with the lens ($z_s=z_l$). To alleviate this effect, we can try to exclude all galaxies that are likely cluster members from the shape catalogue. However, due to intrinsic imperfections in the cuts, some of these galaxies leak into the source catalogue used in the weak lensing measurement. Since foreground and physically associated galaxies are unlensed, the inclusion of these galaxies will cause $\Delta\Sigma$ to be underestimated (the dilution effect). Therefore, the $\Delta\Sigma$ measurements must be boosted to recover the true lensing signal, the so-called \textit{boost-factor} correction \citep{2003ApJ...598..804K, 2004AJ....127.2544S, 2014MNRAS.439...48A, 2015MNRAS.449..685H, 2017MNRAS.466.3103S, 2017MNRAS.467.3024L, 2017MNRAS.469.4899M, 2019MNRAS.482.1352M, 2019MNRAS.489.2511V}. 

\begin{figure}
\includegraphics[width=\linewidth]{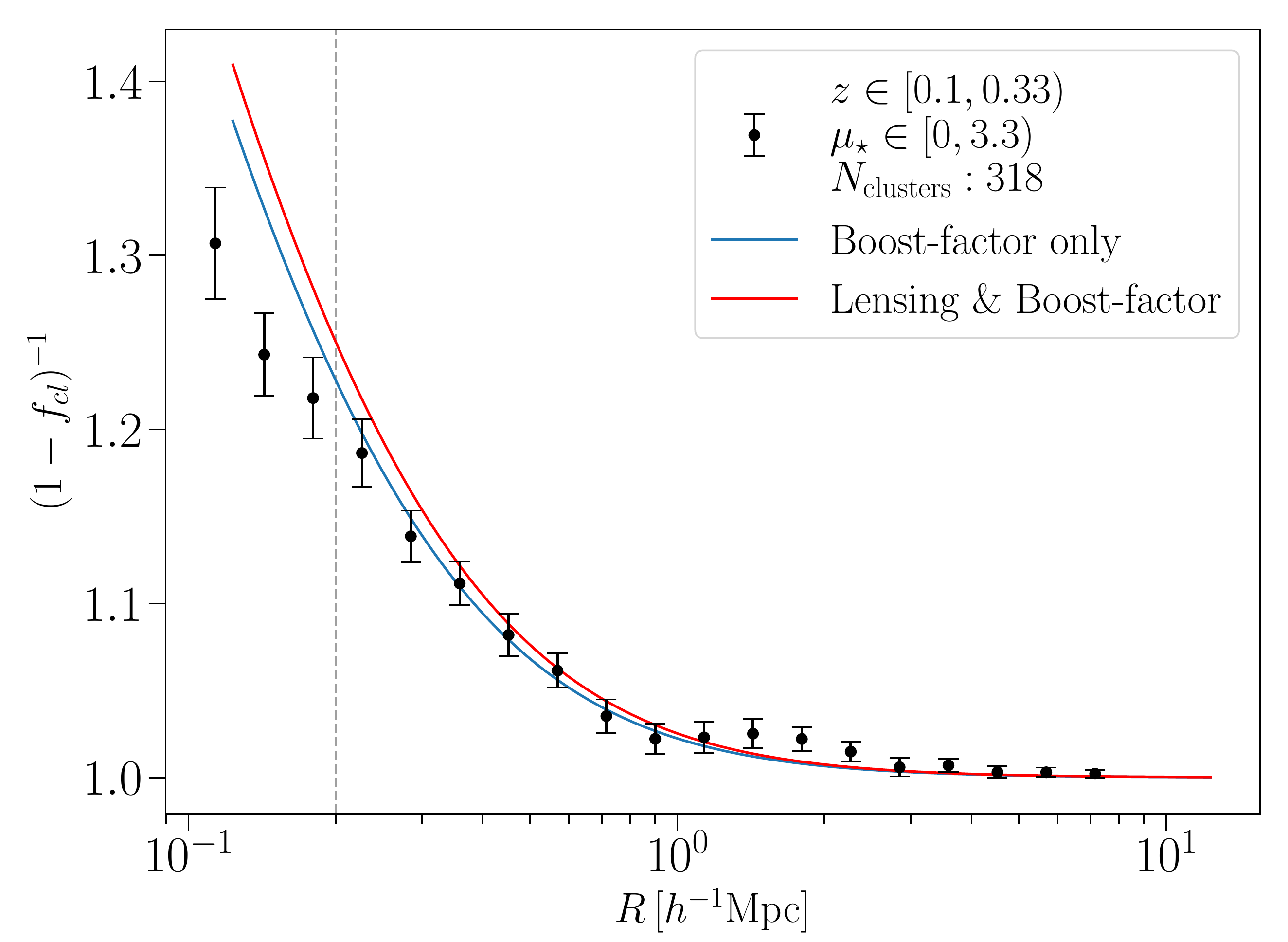}
\caption{Boost-factor measurement of a single profile with $z\in[0.1;0.33)$ and $\mu_{\star}\in[0;3.3)\times 10^{12}\,\mathrm{M_{\odot}}$ (\textit{black dots}). The \textit{blue curve} is the best-fit for fitting the boost-factor data alone and the \textit{red curve} is the best-fit for the joint fit of the lensing and boost-factor data as described in \autoref{sec:full_model}. The vertical dashed line shows the inner radial limit in which we performed our boost-factor fits.}
\label{fig:boostfactor}
\end{figure}

We determine the boost-factor correction by following \cite{2017MNRAS.468..769G, 2017MNRAS.469.4899M, 2019MNRAS.482.1352M, 2019MNRAS.489.2511V} that make use of the estimated $p(z)$ of the source galaxy sample to calculate the cluster contamination fraction $f_{\rm cl}$ and the corresponding covariance matrix $C_{\rm f_{\rm cl}}$ estimated from jackknife resampling. Then, $f_{\rm cl}$ is used to recover the lensing profile corrected from contamination as
\begin{equation}
	\label{eq:boost_correction}
	\widetilde{\Delta\Sigma}_{\rm corr}(R) = \frac{\widetilde{\Delta\Sigma}(R)}{1 - f_{\rm cl}(R)}\,.
\end{equation}  
The $p(z)$ decomposition method for obtaining the boost factor $f_{\rm cl}$ is described in detail and validated on simulated DES-like mock catalogues in \cite{2019MNRAS.489.2511V}. In \autoref{fig:boostfactor} we show an example of the measured boost-factor profile for the stack with $z\in[0.1;0.33)$ and $\mu_{\star}\in[0;3.3)\times 10^{12} \, \mathrm{M_{\odot}} $.     

Following \cite{2019MNRAS.482.1352M} we do not apply \autoref{eq:boost_correction} directly to our data but instead we dilute the amplitude of our model for the predicted profiles. By parameterising the boost-factor as $\mathcal{B} \equiv (1 - f_{\rm cl})^{-1}$ we model the cluster-member contamination by an NFW-like profile, with two free parameters ($B_0$ and $R_s$), in the form  
\begin{equation}
  \label{eq:boost_model}
  {\mathcal B}(R) = 1+B_0\frac{1-F(x)}{x^2-1}\,,
\end{equation}
where $x=R/R_s$, and
\begin{align}
  \label{eq:boost_model2}
  F(x) = \left\{
  \begin{array}{lr}
    \frac{\tan^{-1}\sqrt{x^2-1}}{\sqrt{x^2-1}} & : x > 1\\
    1 & : x = 1\\
    \frac{\tanh^{-1}\sqrt{1-x^2}}{\sqrt{1-x^2}} & : x < 1
  \end{array}
  \right.\,.
\end{align}
The implementation of this model is available in the Python library \texttt{cluster\_toolkit} \href{https://github.com/tmcclintock/cluster_toolkit}{\faGithubAlt}. 

For each stack, we fit the measured boost-factors along with the respective lensing profile, which introduces two additional parameters in our $\Delta\Sigma$ model, the normalisation $B_0$ and the scaling radius $R_s$. When performing this joint fit of the lensing and boost-factor profiles, we try to follow \cite{2019MNRAS.482.1352M} and use their flat priors for the boost-factor parameters. However, this choice leads to ``unrealistic'' values for $B_0$ and $R_s$. For instance, we expect a small value for the scaling radius $(R_s << 1 h^{-1 }\mathrm{Mpc})$, since we have a peak in the contamination fraction at low radii (see Section 4.1 of \cite{2019MNRAS.489.2511V}). However, in the joint fit, we find large values for $R_s$, dominated by the upper limits of our priors. \cite{2019MNRAS.482.1352M} also have shown that $B_0$ and $R_s$ are highly degenerate (see their Figure 10), and this might have an impact in our ability to constrain these parameters when performing the joint fit with lensing using flat priors. Therefore, we decide to perform a separated fit of the boost-factor profiles alone and use the derived values for the parameters (see \autoref{tab:boost_parameter_posteriors}) as input in a Gaussian prior when performing the joint fit with the lensing profiles (see \autoref{sec:full_model}).    
\begin{table}
	\caption{Best-fit parameters from fitting the boost-factors without the lensing profiles. We use the following flat priors: $R_s = [0, 10]$ and $B_0=[0, 1]$.}

\begin{center}
\begin{tabular}{ c|c|c|c } 
 \hline
 $\mu_{\star}\,[10^{12} M_{\odot}]$ & $z$ & $B_0$ & $R_s\,[\mathrm{Mpc}]$ \\ 
 \hline
 \hline
 $[0.0;3.3)$  & \multirow{4}{*}{$[0.1;0.33)$} & $0.57\pm 0.23$ & $0.19\pm 0.06$ \\ 
 $[3.3;5.0)$  &                               & $0.26\pm 0.12$ & $0.46\pm 0.21$ \\ 
 $[5.0;7.5)$  &                               & $0.30\pm 0.10$ & $0.49\pm 0.15$ \\ 
 $[7.5;60.0)$ &                               & $0.20\pm 0.03$ & $1.01\pm 0.17$ \\  
 \hline
 $[0.0;4.04)$  & \multirow{4}{*}{$[0.33;0.5)$} & $0.23\pm 0.34$ & $0.01\pm 0.03$ \\ 
 $[0.04;5.65)$ &                               & $0.22\pm 0.26$ & $0.14\pm 0.11$ \\ 
 $[5.65;8.05)$ &                               & $0.15\pm 0.10$ & $0.23\pm 0.09$ \\ 
 $[8.05;60.0)$ &                               & $0.07\pm 0.02$ & $0.84\pm 0.24$ \\  
 \hline
 $[0.0;3.88)$  & \multirow{4}{*}{$[0.5;0.65)$} & $0.22\pm 0.34$ & $0.02\pm 0.03$ \\ 
 $[3.88;5.42)$ &                               & $0.23\pm 0.34$ & $0.13\pm 0.34$ \\ 
 $[5.42;7.68)$ &                               & $0.08\pm 0.23$ & $0.34\pm 0.65$ \\ 
 $[7.68;60.0)$ &                               & $0.29\pm 0.26$ & $0.16\pm 0.09$ \\  
\end{tabular}
\end{center}
	\label{tab:boost_parameter_posteriors}
\end{table}

\subsubsection{Reduced shear}
\label{sec:red_shear}

In practice, we measure the reduced shear $\mathbf{g}$ instead of the true shear $\gamma$ (see \autoref{eq:redu_shear}). To account for this approximation, we multiply our $\Delta\Sigma$ model by the factor
\begin{equation}
  \label{eq:reduced_shear}
  \mathcal{G}(R) = \frac{1}{1-\kappa} = \frac{1}{1-\Sigma(R)\Sigma_{\rm crit}^{-1}}\,,
\end{equation}
where $\Sigma_{\rm crit}^{-1}$ is defined in \autoref{sigma_crit} and $\Sigma(R)$ is
\begin{equation}
	\label{eq:reduced_shear_sigma}
    \Sigma(R) = p_{\rm cc}\Sigma_{\rm cen} + (1-p_{\rm cc}) \Sigma_{\rm mis}\,,
\end{equation}
where $\Sigma_{\rm cen}$ comes from \autoref{eq:Sigma} and $\Sigma_{\rm mis}$ from \autoref{eq:sigma_mis}. However, this correction is expected to have a negligible effect in our results. 

\subsubsection{Shear and photo-z bias}
\label{sec:shear_photoz}

In the weak lensing analysis, two major sources of systematics are the shape measurements and photo-z uncertainties. The former can lead to wrong shear estimates and the latter can bias the distance measurements leading to a biased $\Sigma_{\rm crit}$, consequently affecting our $\Delta\Sigma$ estimates. \cite{2018MNRAS.481.1149Z} have tested for several sources of bias in the shear measurements, in particular, self-calibration of the images allowed them to determine the multiplicative $m$ and the additive $c$ biases. They found no evidence of a significant additive bias term but estimated the multiplicative bias to be $m = 0.012 \pm 0.013$. 

\cite{2018MNRAS.478..592H} and \cite{2019MNRAS.482.1352M} present a method to calibrate the photo-z estimates with precise measurements from COSMOS bands to determine the bias and its uncertainties. Briefly, they match the DES lensing source galaxies and the COSMOS galaxies according to their flux in each band and their intrinsic size. Following the same selection and weight as in \autoref{projected-ds}, we compute the true weighted mean $\Sigma_{\rm crit, TRUE}^{\prime\, -1}$ from the matched COSMOS sample. The MOF $griz$ BPZ redshift distribution samples provide a mean $\Sigma_{\rm crit, MEAS}^{\prime\, -1}$ that connects the weighted mean tangential shear to the $\Delta\Sigma$ profile. Since the source selection for these measurements depends on the lens redshift, we need to repeat them in the cluster redshift range sampled in our analysis $z_l=0.1-0.65$.            

\begin{figure}
\includegraphics[width=\columnwidth]{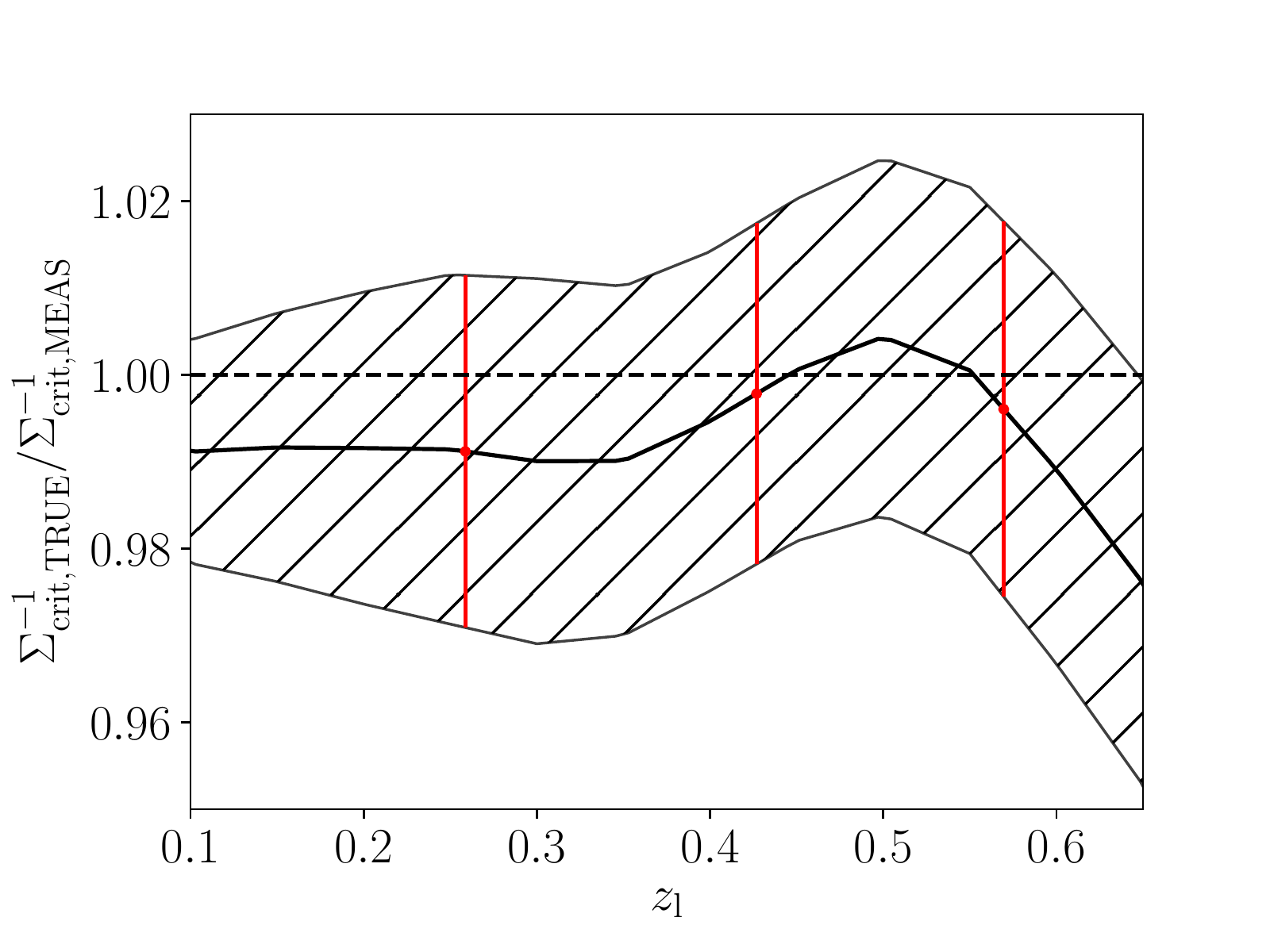}
\caption{The photo-z correction factor to $\Sigma_{\rm crit}^{-1}$ as described in \autoref{sec:shear_photoz}. The \textit{gray hatched} region indicates the $1\sigma$ range of the correction factor. \textit{Red points} with error bars show the correction factors applied in each redshift bin.}
\label{fig:photozbias}
\end{figure}

Following \cite{2019MNRAS.482.1352M}, the model for the bias takes into account four sources of uncertainty in the calibration of photometric redshift distributions: $i)$ cosmic variance; $ii)$ photometric zeropoint offsets; $iii)$ morphology matching; $iv)$ systematic uncertainty of the matching algorithm, and it is given by \begin{equation}
	\label{eq:sigmacrit_ratio}
	\frac{\Sigma_{\rm crit, MEAS}^{'-1}}{\Sigma_{\rm crit, TRUE}^{'-1}} \equiv 1+\delta\, ,
\end{equation}
where the quantity $\delta$ is the offset between the true mean inverse critical surface density from COSMOS and $\Sigma_{\rm crit}^{'-1}$ from our photo-z estimates. We show in \autoref{fig:photozbias} the dependence of this ratio on lens redshift in the range of our analysis. 

We incorporate $\delta$ in our analysis as a prior that varies between each stack. The variation across the cluster redshift bins are
\begin{equation}
	\label{eq:multiplicative_photoz_bias}
	\delta = \begin{cases}
		0.009 \pm 0.021 & {\rm for}\ z\in[0.1,0.33) \\
		0.002 \pm 0.020 & {\rm for}\ z\in[0.33,0.5) \\
		0.004 \pm 0.022 & {\rm for}\ z\in[0.5,0.65).
	\end{cases}
\end{equation}

We combine the shear and photo-z bias ($m$ and $\delta$) to define the factor $\mathcal{A}_m = 1 + m + \delta$, which is included in the final likelihood as the prior
\begin{equation}
	\label{eq:multiplicative_total_bias}
	\mathcal{A}_m = \begin{cases}
		1.021 \pm 0.024 & {\rm for}\ z\in[0.1,0.33) \\
		1.014 \pm 0.024 & {\rm for}\ z\in[0.33,0.5) \\
		1.016 \pm 0.025 & {\rm for}\ z\in[0.5,0.65).
	\end{cases}
\end{equation}

\subsubsection{Triaxiality and projection effects}
\label{sec:triax_proj}

Cluster finders that rely on photometric data to identify galaxy clusters typically select systems that are aligned along the line-of-sight with higher probability. The photometric cluster selection can also be affected by the presence of other objects along the line-of-sight. We refer to these two effects as \textit{triaxiality} and \textit{projection effects}, which both affect the measured cluster MOR \citep{2011ApJ...728..126W, 2012MNRAS.426.2046A, 2012MNRAS.426.1829N, 2014MNRAS.443.1713D}.

\cite{2017MNRAS.469.4899M} determined the projection effect correction factor by modelling the ratio between the average cluster stack mass not affected by projections $\langle M \rangle_0$ and the average mass $\langle M \rangle$ of the cluster affected by projections. They model the projected cluster as a sum of a primary halo that must have at least a mass  $0.5 \langle M \rangle_0$ and an excess mass $\epsilon \langle M \rangle_0$, where $\epsilon \in [0.0, 0.5]$. Then, for a fraction $p$ of clusters affected by projections, they write the average mass of the cluster stack as
\begin{equation}
\langle M \rangle = (1-p) \langle M \rangle_0 + p(0.5 + \epsilon) \langle M \rangle_0.    
\end{equation}
To recover the mass in the absence of projections, i. e. $\langle M \rangle_0$, we should multiply the recovered weak lensing masses by
\begin{equation}
\frac{\langle M \rangle_0}{\langle M \rangle} = \frac{1}{1+p(\epsilon - 0.5)} = 1.02 \pm 0.02\, ,     
\end{equation}
where the numerical value above was estimated from $10^4$ Monte Carlo realisations of $p$ and $\epsilon$. They adopted a Gaussian prior for $\epsilon$ of $\epsilon=0.25 \pm 0.15$ such that $\epsilon=0$ and  $\epsilon=0.5$ are within $2\sigma$ of the central value, and $p = 10\% \pm 4\%$ as estimated from \cite{2017MNRAS.466.3103S}. In our analysis we are using the same clusters as \cite{2019MNRAS.482.1352M}, i. e. a richness-selected sample, therefore we believe that it is reasonable to apply the same correction for the projection effects.   

Using simulation of richness-selected clusters, \cite{2014MNRAS.443.1713D} estimated that triaxiality can overestimate cluster masses by $4.5\% \pm 1.5\%$. \cite{2017MNRAS.469.4899M} argued that this estimate can be understood as correlated scatter between richness and weak lensing masses leading weak lensing masses to overestimate cluster masses by an amount of $\exp(-\beta r \sigma_{ \ln M | \lambda } \sigma_{\ln M | M_{\mathrm{WL}}})$, where $\beta$ is the slope of the halo mass function, $r$ is the correlation coefficient between richness and weak lensing mass and $\sigma$'s are the intrinsic scatters in the correspondent scaling relations. \cite{2017MNRAS.469.4899M} adopted $r \in [0, 0.5]$ \citep{2012MNRAS.426.1829N}, $\sigma_{\ln M | \lambda} = 0.25 \pm 0.05$ \citep{2014ApJ...783...80R}, $\sigma_{\ln M| M_{\mathrm{WL}}} = 0.25 \pm 0.05$ and $\beta \sim 3$ to arrive at a correction factor of $0.96 \pm 0.02$. \cite{2020MNRAS.493.4591P} found that $\sigma_{\ln M |\mu_{\star}} = 0.26^{+0.15}_{-0.10}$, then we could argue that we would arrive at a similar correction factor for triaxiality as in \cite{2017MNRAS.469.4899M} for our $\mu_{\star}$ sample.           

Therefore, we follow \cite{2017MNRAS.469.4899M} and \cite{2019MNRAS.482.1352M} and apply their corrections to triaxiality and projection effects. For triaxiality we use a multiplicative factor given by the Gaussian $\mathcal{G}(0.96,0.02)$ and for projection effects a factor $\mathcal{G}(1.02,0.02)$. In both cases we use a random draw to determine the multiplicative factors to be applied in the masses together with the model bias correction, which will be described in \autoref{sec:mod_sys}.

While this analysis was in internal review by the collaboration, the cluster cosmology results from DES Y1 was released \citep{2020arXiv200211124D} and they found that the $\sigma_8 - \Omega_m$ posteriors are in $2.4\sigma$ tension with DES Y1 3x2pt analysis, and in $5.6\sigma$ with Planck CMB results. They argue that this tension is most likely driven by systematics in the weak-lensing mass calibration that were not fully modelled. Currently, photometric redshifts together with triaxiality and projection effects are the systematics with the largest contributions to the error budget in the mass-calibration with richness \citep{2019MNRAS.482.1352M}. However, none of these systematics alone were found to explain the tension in the DES Y1 cluster cosmology result. But it was shown that the proposed projections and triaxiality corrections applied in \cite{2017MNRAS.469.4899M, 2019MNRAS.482.1352M} are probably not enough (see Figure 12 in \citealt{2020arXiv200211124D}) for our current measurements, in particular, for clusters with $\lambda \in (20,30]$. Therefore, we need to improve our understanding of the low richness cluster sample to find a better model for projections and triaxiality, both in a mass-richness and in a mass-$\mu_{\star}$ calibration analysis. Since this is beyond the scope of this paper, then, we present our results with the corrections described in this section, acknowledging that we may not be fully accounting for the projection and triaxiality effects in our mass estimates.                       

\subsection{The full model}
\label{sec:full_model}

The multiplicative corrections described in the previous sections are combined with our full model of the weak lensing profile in the form
\begin{equation}
	\label{eq:full_delta_sigma_model}
	\Delta\Sigma = \frac{\mathcal{A}_{\rm m}\mathcal{G}(R)}{\mathcal{B}(r)}\left[p_{cc}\Delta\Sigma_{\rm NFW} + (1 - p_{cc})\Delta\Sigma_{\rm misc}\right]\,.
\end{equation}
This model includes the multiplicative bias $\mathcal{A}_{\rm m}$, the boost factor $\mathcal{B}(r)$, the reduced shear correction $\mathcal{G}(R)$ and the miscentring parameter $p_{\rm cc}$. The log-likelihood of the $k$th $\Delta\Sigma$ profile is
\begin{equation}
	\label{eq:deltasigma_likelihood}
    \ln \lkhd(\Delta\Sigma_k\, | \, M_k,p_{\rm cc},\mathcal{A}_{\rm m}, B_{\rm 0},R_{\rm s}) \propto -\frac{1}{2}{\rm \bf D}_k^T{\rm C}^{-1}_{\Delta\Sigma}{\rm \bf D}_k
\end{equation}
where ${\rm \bf D} = (\widetilde{\Delta\Sigma} - \Delta\Sigma)_k$ with $\widetilde{\Delta\Sigma}$ computed from \autoref{eq:updated_deltasigma_estimate} and $C_{\Delta\Sigma}$ is the jackknife covariance matrix of $\Delta\Sigma$. The corresponding log-likelihood of the measured $f_{{\rm cl},k}$ in the $k$th cluster subset given the parameters in \autoref{eq:boost_model} is 
\begin{equation}
  \label{eq:boost_likelihood}
  \ln \lkhd(f_{{\rm cl},k}\,|\, B_0, R_s) \propto -\frac{1}{2}{\rm \bf B}_k^T{\rm C}^{-1}_{f_{\rm cl}}{\rm \bf B}_k,
\end{equation}
where ${\rm \bf B}_k = (\mathcal{B}-\mathcal{B}_{\rm model})_k$ and $C_{f_{\rm cl}}$ is the covariance matrix of the boost-factor, also obtained from jackknife.

The weak lensing and boost-factor profiles are fitted simultaneously with the total log-likelihood for a single cluster subset computed as
\begin{equation}
	\label{eq:total_likelihood}
    \begin{split}
		\ln \lkhd_k =& \ln \lkhd(\Delta\Sigma_k\,|\, M_k,p_{\rm cc},\mathcal{A}_{\rm m}, B_{\rm 0},R_{\rm s}) +\\
    		&\ln \lkhd(f_{{\rm cl},k}\,|\, B_{\rm 0}, R_{\rm s})\,.
    \end{split}
\end{equation}
Note that while the fit of $\Delta\Sigma$ and boost-factor is performed in conjunction, each cluster subset is fitted independently of the other subsets. Also, note that in our approach, the constraints on the boost-factor parameters are informed by both their dilution effect on the $\Delta\Sigma$ profile (as shown in \autoref{eq:full_delta_sigma_model}) as well as independent measurements of $f_{\rm cl}$ (see an example of such measurement in \autoref{fig:boostfactor}).      

\begin{table}
	\setlength{\tabcolsep}{.4em}
	\caption{Parameters in the lensing likelihood $\mathcal{L}(\Delta\Sigma)$ (\autoref{eq:deltasigma_likelihood}) and boost-factor likelihood $\mathcal{L}(\mathcal B)$ (\autoref{eq:boost_likelihood}). Flat priors are specified with limits in square brackets, Gaussian priors with means $\pm$ standard deviations.}
	\begin{tabular}{lll}
		Parameter & Description & Prior \\ \hline
		$M_{\rm 200c}$ & Halo mass & $[10^{11},10^{18}]$ \\
		$p_{\rm cc}$ & Correctly centred fraction & $0.75 \pm 0.08$ \\ 
		$A_{m}$&Shape \& \photoz\ bias & \autoref{eq:multiplicative_total_bias}\\
		$B_0$ & Boost magnitude & \autoref{tab:boost_parameter_posteriors} \\
		$R_s$ & Boost factor scale radius & \autoref{tab:boost_parameter_posteriors} \\
	\end{tabular}
    \label{tab:modeling_parameters}
\end{table}

A list of the model parameters describing each cluster stack and their corresponding priors are summarised in \autoref{tab:modeling_parameters}. We use the Bayesian formalism and the Monte Carlo Markov Chain (MCMC) method through the package \texttt{emcee} \href{https://github.com/dfm/emcee}{\faGithubAlt} \citep{2013PASP..125..306F} to perform the likelihood sampling. We use 64 walkers with 10000 steps each, discarding the first 2000 steps of each walker as burn-in. We also verify the autocorrelation time of the chains to check their convergence. To avoid confirmation bias, we blind the chains before applying the corrections of triaxiality and projections effects and model bias, which we will describe in the next section. Our blinding procedure relies on randomly shifting the peak of the posterior distribution of $M_{\rm 200c}$ in the chains. 

\subsection{Modelling systematics}
\label{sec:mod_sys}

The analytical model for the centred term, $\Delta\Sigma_{\mathrm{NFW}}$ \citep{1996A&A...313..697B, 1999astro.ph..8213O}, can present differences from the true $\Delta\Sigma$ profiles of the cluster halos of mean mass $M$. These deviations are due to the mismatch of density profiles in simulations \citep{2017MNRAS.469.4899M, 2018ApJ...854..120M, 2019MNRAS.482.1352M}, in particular, in the transition between the one and two halo regimes, which can bias the recovered weak lensing masses. Therefore, we need to calibrate our model with simulations.  

In order to achieve that, we measure the weak-lensing masses of dark matter halos in $N$-body simulations using the same formalism we employ to the DES data. The halos are drawn from an $N$-body simulation of a flat $\Lambda$CDM cosmology run with \textsc{Gadget} \citep{2005MNRAS.364.1105S}. The simulation uses 1400$^3$ particles in a box with $1050\ h^{-1}{\rm Mpc}$ on a side with periodic boundary conditions. The force softening is $20\ h^{-1}{\rm kpc}$. The simulation was run with the cosmology $\Omega_m=0.318$, $h=0.6704$, $\Omega_b=0.049$, $\tau=0.08$, $n_s=0.962$, and $\sigma_8=0.835$. Halos of mass $10^{13}\ h^{-1}{\rm M}_\odot$ are resolved with 100 particles. Halos are defined using a spherical overdensity mass definition of 200 times the background density and are identified with the \textsc{ROCKSTAR} halo finder \citep{2013ApJ...762..109B}.

The simulation is used to construct the synthetic $\Delta\Sigma$ profiles of halos at four different snapshots: $z\in[0,0.25,0.5,1]$. We assigned a $\mu_{\star}$ to each halo by inverting the mass--$\mu_{\star}$ relation of \citet{2018MNRAS.474.1361P} and adding 25 per cent scatter. Then, we grouped our halos into $(z, \mu_{\star})$ subsets identical to how we grouped our real clusters. For each of these halo subsets we measured the halo-matter correlation function with the \citet{1993ApJ...412...64L} estimator implemented in \texttt{Corrfunc} \href{https://github.com/manodeep/Corrfunc}{\faGithubAlt} code \citep{2017ascl.soft03003S}. We numerically integrate the halo-matter correlation function to obtain the $\Delta\Sigma$ profile. The resulting simulated $\Delta\Sigma$ profile is a combination of the $\Delta\Sigma_{\mathrm{NFW}}$ and a 2-halo term. 

Note that this $\Delta\Sigma$ profile does not contain any of the systematics that exists in the real data. To incorporate the systematics, we modify the simulated $\Delta\Sigma$ profiles by applying the corrections in \autoref{eq:full_delta_sigma_model}. The miscentring profile $\Delta\Sigma_{\mathrm{misc}}$ is computed by providing as input the true mass from the simulation and the miscentring distribution discussed in \autoref{sec:misc_corr}. The values of $p_{\mathrm{cc}}$ and $\mathcal{A}_\mathrm{m}$ are the central values described in \autoref{tab:modeling_parameters}. For the boost-factor correction $\mathcal{B}(R)$ described in \autoref{sec:mod_boost}, the values for $B_0$ and $R_s$ are obtained from modelling the boost-factors data independently. To apply the reduced shear correction $\mathcal{G}(R)$ described in \autoref{sec:red_shear} in the simulation we use the same $\Sigma_{\mathrm{crit}}^{-1}$ of the real data. Note that in the real data we just have three bins of redshifts, therefore we repeat the values of the third $z$-bin for the snapshot with $z=1$ in the simulations.              

We obtain the observed mass $M_{\mathrm{obs}}$ for this simulated profile by using the same pipeline we apply on the real data, restricting ourselves to the same radial scales employed in the weak lensing analysis, and utilising the covariance matrices recovered from the data to ensure that the simulated data are weighted in the same way as the observed data.

\begin{figure}
\includegraphics[width=\linewidth]{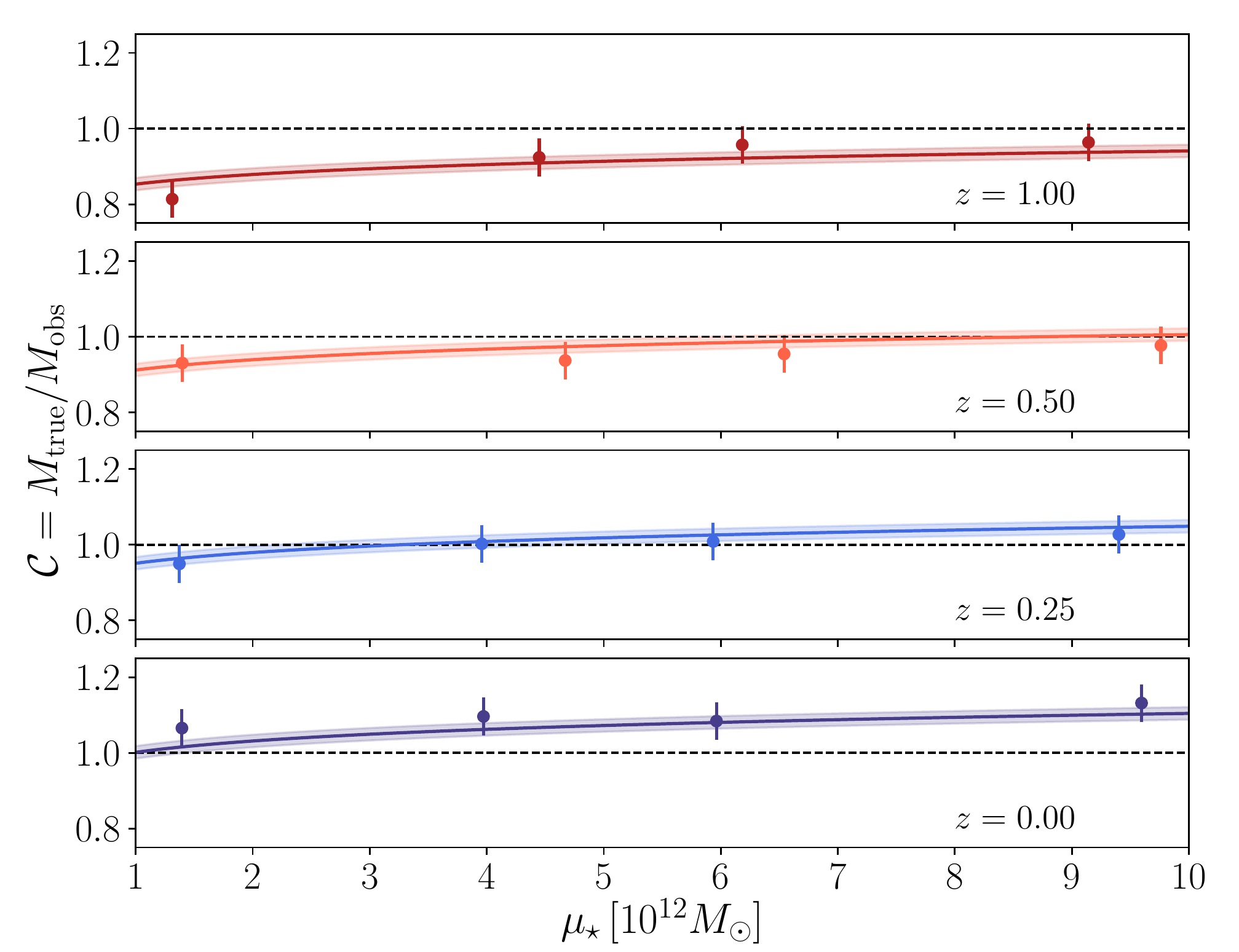}
\caption{The mass bias calibration $\mathcal{C} = M_{\rm true}/M_{\rm obs}$ caused by the adopted analytical form of the $\Delta\Sigma$ profile, as a function of $\mu_{\star}$ for each simulated redshift bin. The solid line and shaded region are the best-fit bias model from \autoref{eq:calibration_model} and 1$\sigma$ uncertainty of the calibration, respectively. Error bars on the measured calibrations are the fitted intrinsic scatter $\sigma_{\mathcal{C}}$. Here we multiply the error bars by a factor of 3 for a better visualisation.}
\label{fig:calibration}
\end{figure}

Defining $M_{\rm true}$ as the mean mass of the halos in the simulated stack, the calibration for each simulated profile is shown in \autoref{fig:calibration}. The model bias calibration $\mathcal{C}=M_{\rm true}/M_{\rm obs}$ was modelled as a function of the mean $\overline{\mu_{\star}}$ and redshift snapshot $z$ of the simulated stack as
\begin{equation}
	\label{eq:calibration_model}
    \mathcal{C}(\overline{\mu_{\star}},z) = C_0\left(\frac{\overline{\mu_{\star}}}{5.16\times 10^{12} M_{\odot}}\right)^\alpha \left(\frac{1+z}{1+z_0}\right)^\beta,
\end{equation}
with $z_0=0.5$ as pivot redshift. The free parameters in the fit are $C_0$, $\alpha$, $\beta$ and the intrinsic scatter $\sigma_{\mathcal{C}}$ of the calibration, and they are determined via a Bayesian fit using flat priors for the parameters (see \autoref{tab:modelbias_parameters}). 

\begin{table}
	\setlength{\tabcolsep}{.4em}
	\caption{Parameters in the model bias fit. Flat priors are specified with limits in square brackets.}
	\begin{tabular}{lll}
		Parameter & Description & Prior \\ \hline
		$C_0$ & Normalisation & $[0, \infty]$ \\
		$\alpha$ & Slope in $\mu_{\star}$ & $[-10,10]$ \\ 
		$ \beta$ & Slope in $z$ & $[-10,10]$ \\
		$\ln\, (\sigma_{\mathcal{C}}^2)$ & Intrinsic scatter & $[-10,10]$\\
	\end{tabular}
    \label{tab:modelbias_parameters}
\end{table}

\begin{figure}
\includegraphics[width=\linewidth]{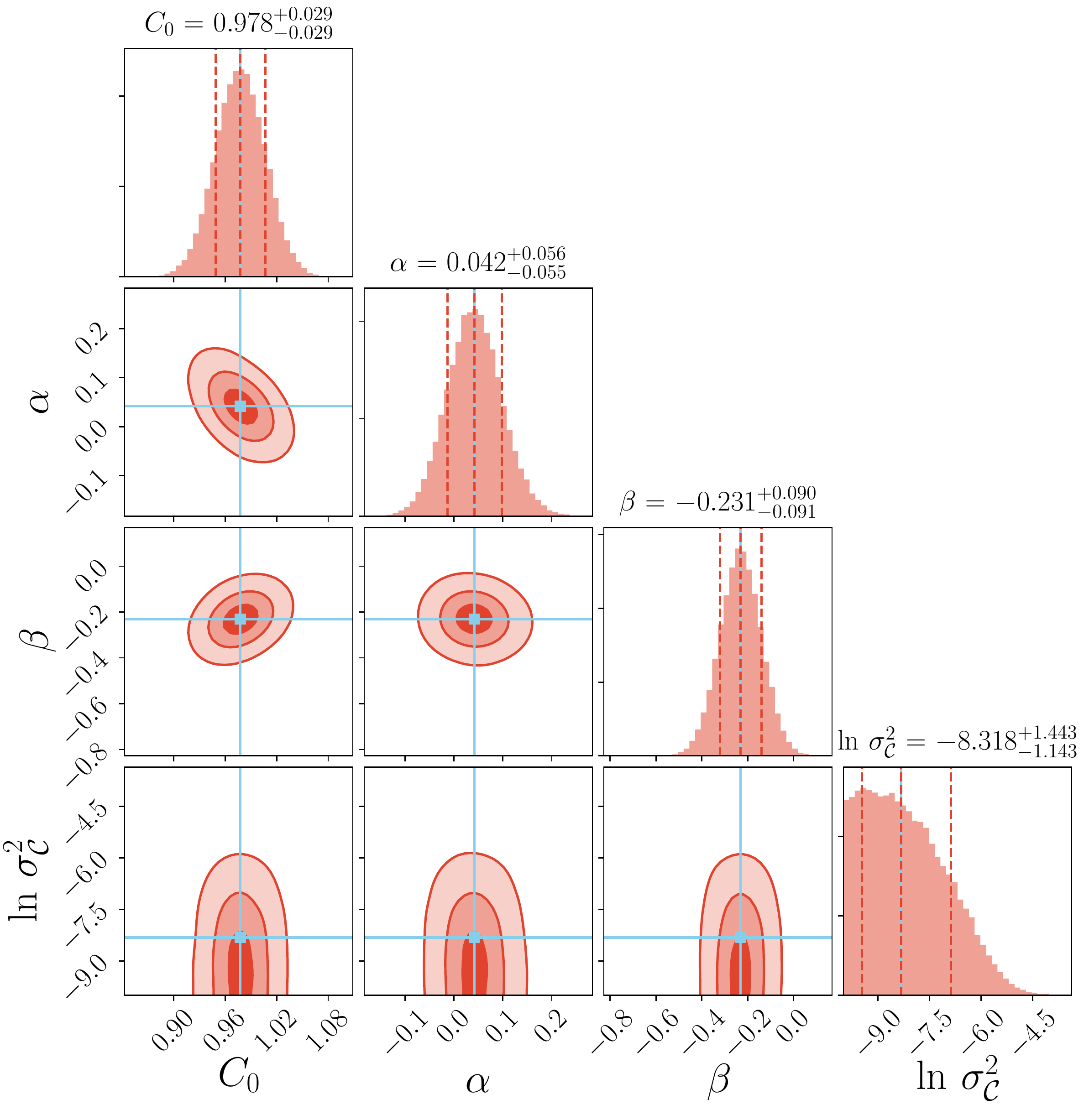}
\caption{Parameters of the $\mathcal{C}(\overline{\mu_{\star}},z)$ relation in \autoref{eq:calibration_model}. Contours are the 1$\sigma$, 2$\sigma$ and 3$\sigma$ confidence areas.}
\label{fig:corner_modelcalibration}
\end{figure}

The mean model bias for our simulated stacks is $\sim 5$ per cent with $C_0 = 0.978 \pm 0.029$, $\alpha = 0.042 \pm 0.055$, $\beta = -0.231 \pm 0.090$ and intrinsic scatter $\sigma_{\mathcal{C}} = 0.016$. In \autoref{fig:corner_modelcalibration} we show the contour plots for these parameters.

We repeated this analysis for profiles assuming different amounts of intrinsic scatter in the $M-\mu_{\star}$ relation from 10 per cent up to 45 per cent. We found that the amount of model bias does not present a significant change with scatter in the $M-\mu_{\star}$ relation. We also have checked that the model bias has negligible changes when we consider the effect of selecting in $\lambda$ but binning in $\mu_{\star}$. To mimic this effect in the simulations we populated the halos with $\lambda$ and $\mu_{\star}$, by inverting the $M-\lambda$ relation from \cite{2017MNRAS.469.4899M} and the $M-\mu_{\star}$ relation from \cite{2018MNRAS.474.1361P}, with scatters $\sigma_{M|\lambda}$ and $\sigma_{M|\mu_{\star}}$. Then, we selected halos with $\lambda > 20$ and stacked them in bins of $z$ and $\mu_{\star}$, verifying the change in the model bias with the different scatters in the $M-\mu_{\star}$ relation. No relevant changes was found. So, we concluded neither the selection effect of selecting in $\lambda$ but ranking in $\mu_{\star}$ nor the different amount of scatters have significant impact in our model bias correction.

In our analysis we focus on the modelling of the centred term by choosing a radial cut, $R =(0.2-2.5)\,\, \mathrm{Mpc}$, to avoid the 2-halo term and we use an analytical expression for the NFW profile. However, the simulated profiles were generated by an integration of the halo-matter correlation function that accounts both for the centred and 2-halo contributions. This might be causing a significant model bias between our model and the simulated profiles, in particular, at the high redshifts snapshots $z=1$ as seen in \autoref{fig:calibration}. Besides that, we are not considering the effect of baryonic physics that might have some impact, especially in the central regions of the halo. Once we apply a radial cut in the inner profile before performing the fit, we expect to minimise the impact of baryonic effects as argued in \cite{2019MNRAS.482.1352M}. However, this will not be true for halos that are miscentred by a large amount.

After this model calibration, the correction for the mean weak lensing mass of a cluster stack in a given bin of $\mu_{\star}$ at redshift $z$ should be
\begin{equation}
\label{m_prime}
M^{\prime} = \mathcal{C}(\mu_{\star}, z) M_{\rm WL},    
\end{equation}
where $M_{\rm WL}$ is the uncalibrated mass estimate and the calibration factor $\mathcal{C}(\mu_{\star}, z)$ is determined by randomly picking one value from the simulation posteriors. 

The final posteriors for our weak lensing masses are properly marginalised over the uncertainty in the calibration factor $\mathcal{C}$ as well as triaxiality and projection effects such that 
\begin{equation}
\label{eq:final_cal_mass}
M = \mathcal{G}(0.96,0.02) \times \mathcal{G}(1.02,0.02) \times M^{\prime} \end{equation}
after fitting the lensing and boost-factor data but before modelling the mass--$\mu_{\star}$--$z$ relation. The final unblinded and corrected masses are shown in \autoref{tb:mass_wl_blind}. 

\begin{table*}
\caption{Best-fitting results for redMaPPer clusters in \autoref{fig:des_ds}. In the fitting, we use a concentration-mass relation from \protect\cite{Diemer_2019} to fix $c_{\mathrm{200c}}$ and we also fix the width of miscentring distribution as $\sigma_{\mathrm{off}}=0.133\, h^{-1}$ Mpc. Our final model has five free parameters, the mass $M_{\mathrm{200c}}$, the fraction of clusters that is correctly centred $p_{\mathrm{cc}}$, the shear$+$photo-z bias $\mathcal{A}_{\mathrm{m}}$ and the boost-factor parameters $B_{\mathrm{0}}$ and $R_{\mathrm{s}}$. The weak lensing and boost-factor profiles were fitted simultaneously, but each cluster subset is fitted independently of each other. We note that the posteriors of $p_{\mathrm{cc}}$ is dominated by the priors and one could say that the measurement is non-informative. However, since we are using the values determined by a previous measurement of the corrected centred redMaPPer cluster in comparison to a X-ray sample (\citealt{2019MNRAS.tmp.1291Z}, von der Linden et al., in preparation), we believe that it is reasonable to assume that the used informative prior in $p_{\mathrm{cc}}$ will not bias the recovered masses. For convenience, we also present here the estimated masses converted to the definition $M_{\mathrm{200m}}$ by using the \texttt{Colossus} code.}
\begin{center}
\begin{tabular}{ c|c|c|c|c|c|c|c } 
 \hline
 $\mu_{\star}\,[10^{12} M_{\odot}]$ & $z$ & $M_{\mathrm{200c}}\,[10^{14} \, \mathrm{h^{-1} M_{\odot}}]$ & $M_{\mathrm{200m}}\,[10^{14} \, \mathrm{h^{-1} M_{\odot}}]$ & $p_{\mathrm{cc}}$ & $\mathcal{A}_{\mathrm{m}}$ & $B_0$ & $R_\mathrm{s}\,[\mathrm{Mpc}]$ \\ 
 \hline
 \hline
 $[0.0;3.3)$  & \multirow{4}{*}{$[0.1;0.33)$} & $0.70\pm 0.06$ & $0.90\pm 0.08$ & $0.73\pm 0.08$ & $1.019\pm 0.024$ & $0.59\pm 0.21$ & $0.20\pm 0.06$ \\ 
 $[3.3;5.0)$  &                               & $1.18\pm 0.11$ & $1.52\pm 0.14$ & $0.72\pm 0.08$ & $1.017\pm 0.023$ & $0.36\pm 0.11$ & $0.61\pm 0.18$ \\ 
 $[5.0;7.5)$  &                               & $1.38\pm 0.12$ & $1.77\pm 0.15$ & $0.69\pm 0.08$ & $1.016\pm 0.024$ & $0.40\pm 0.08$ & $0.63\pm 0.13$ \\ 
 $[7.5;60.0)$ &                               & $2.46\pm 0.16$ & $3.16\pm 0.20$ & $0.64\pm 0.07$ & $1.009\pm 0.024$ & $0.24\pm 0.03$ & $1.21\pm 0.16$ \\  
 \hline
 $[0.0;4.04)$  & \multirow{4}{*}{$[0.33;0.5)$} & $0.81\pm 0.09$ & $0.98\pm 0.11$ & $0.77\pm 0.08$ & $1.015\pm 0.023$ & $0.33\pm 0.25$ & $0.02\pm 0.02$ \\ 
 $[0.04;5.65)$ &                               & $0.89\pm 0.10$ & $1.07\pm 0.12$ & $0.76\pm 0.08$ & $1.015\pm 0.023$ & $0.26\pm 0.20$ & $0.14\pm 0.09$ \\ 
 $[5.65;8.05)$ &                               & $1.34\pm 0.12$ & $1.64\pm 0.14$ & $0.69\pm 0.07$ & $1.011\pm 0.023$ & $0.21\pm 0.09$ & $0.27\pm 0.09$ \\ 
 $[8.05;60.0)$ &                               & $2.24\pm 0.14$ & $2.73\pm 0.17$ & $0.74\pm 0.07$ & $1.012\pm 0.023$ & $0.07\pm 0.02$ & $0.86\pm 0.23$ \\  
 \hline
 $[0.0;3.88)$  & \multirow{4}{*}{$[0.5;0.65)$} & $0.66\pm 0.12$ & $0.77\pm 0.14$ & $0.75\pm 0.08$ & $1.015\pm 0.025$ & $0.33\pm 0.25$ & $0.03\pm 0.02$ \\ 
 $[3.88;5.42)$ &                               & $0.83\pm 0.12$ & $0.97\pm 0.14$ & $0.74\pm 0.08$ & $1.016\pm 0.025$ & $0.26\pm 0.22$ & $0.12\pm 0.09$ \\ 
 $[5.42;7.68)$ &                               & $1.09\pm 0.14$ & $1.28\pm 0.16$ & $0.74\pm 0.08$ & $1.016\pm 0.025$ & $0.09\pm 0.10$ & $0.33\pm 0.25$ \\ 
 $[7.68;60.0)$ &                               & $1.91\pm 0.17$ & $2.24\pm 0.19$ & $0.73\pm 0.08$ & $1.014\pm 0.025$ & $0.29\pm 0.15$ & $0.16\pm 0.05$ \\  
\end{tabular}
\end{center}
\label{tb:mass_wl_blind}
\end{table*}

\section{The mass--$\mu_{\star}$--redshift relation}
\label{sec:masscal}

We obtain a mass calibration for the galaxy cluster stacks using their weak lensing masses shown in \autoref{tb:mass_wl_blind}. We characterise the mass--$\mu_{\star}$--redshift relation of these clusters as
\begin{equation}
\label{eq:mass_richness_redshift}
\langle M| \mu_{\star},z \rangle = M_{0} \left( \frac{\mu_{\star}}{\mu_{\star}^0} \right)^{F_{\mu_{\star}}} \left(\frac{1+z}{1+z_0}\right)^{G_{z}},    
\end{equation}
where $M_0$, $F_{\mu_{\star}}$ and $G_{z}$ are the free parameters of our model with pivot values $\mu_{\star}^0=5.16\times 10^{12} \mathrm{M_{\odot}}$ and $z_0=0.35$. We model the likelihood for our model as
\begin{equation}
  \label{eq:mass_richness_likelihood}
  \ln \mathcal{L}(M_{\rm obs}\,|\,M_0,F_{\mu_{\star}},G_z) \propto - \frac{1}{2} (\Delta M)^T\ \mathsf{C}_M^{-1}\ (\Delta M)\,,
\end{equation}
where $\Delta M = M - \langle M|\mu_{\star},z \rangle$. Here, the mass $M$ is the value after unblinding and applying the  correction in \autoref{eq:final_cal_mass}. $\mathsf{C}_M$ is the covariance matrix between the mass bins obtained following Section 6.2 of \cite{2019MNRAS.482.1352M}. 

Briefly, to construct the mass covariance, we combine the errors in the mass obtained by performing the profile-fitting in three configurations: $i)$ our fiducial run called \verb|Full|, where we vary all the 5 parameters of our lensing likelihood using the priors in \autoref{tab:modeling_parameters}, and for which the posteriors are reported in \autoref{tb:mass_wl_blind}; $ii)$ \verb|FixAm|, where the shear+photo-z parameter $\mathcal{A}_{\mathrm{m}}$ is fixed to 1 and the other four parameters are free; $iii)$ \verb|OnlyM|, where the only free parameter is the mass. We do not report the posteriors for \verb|FixAm| or \verb|OnlyM| configurations.   

Besides being used for constructing the full mass covariance, we can perform the mass calibration for each of these configurations and use the estimated uncertainties in each parameter to determine the statistical and systematic uncertainties in our final mass calibration. The statistical error is computed by the difference in the variance of the parameters obtained with the masses from \verb|Full| and \verb|FixAm| configurations. The systematic uncertainties are obtained by the difference in the parameter's variance between the \verb|Full| and \verb|OnlyM|.         

\begin{table}
\setlength{\tabcolsep}{.6em}
	\caption{Parameters of the $M$--$\mu_{\star}$--$z$ relation from \autoref{eq:mass_richness_likelihood} with their posteriors. The mass is defined as $M_{200{\rm c}}$ in units of $\mathrm{h}^{-1} {\rm M_{\odot}}$. The pivot $\mu_{\star}$ and pivot redshift correspond to the median values of the cluster sample. Flat priors are specified with limits in square brackets. Uncertainties are the $1\sigma$ confidence intervals and are split into statistical (first) and systematic (second). The posterior of $M_0$ is in units of $\mathrm{h^{-1} M_{\odot}}$.}
	\label{tab:mass_richness_parameters}
    \begin{tabular}{lllll}
    	Parameter & Description & Prior & Posterior \\ \hline
        $M_0$ & Mass pivot & [$10^{11}$, $10^{18}$] & $1.14 \pm 0.05 \pm 0.05 $ \\ 
		$F_{\mu_{\star}}$ & Mass proxy scaling & [-10, 10] & $0.76 \pm 0.01 \pm 0.06 $\\
		$G_z$ & Redshift scaling & [-10, 10] & $-1.14 \pm 0.04 \pm 0.38 $\\
    \end{tabular}
\end{table}

\begin{figure}
  \includegraphics[width=\linewidth]{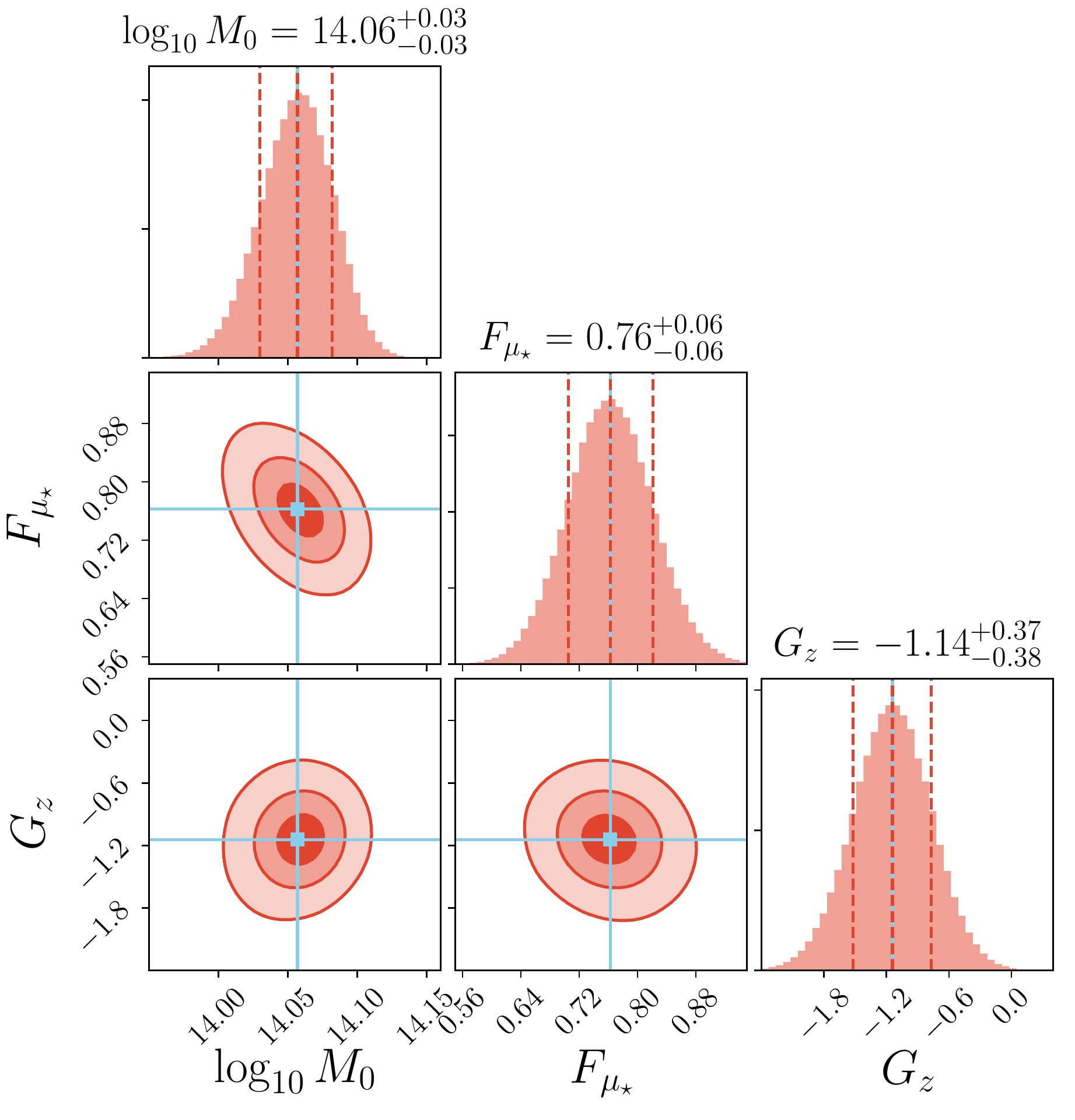}
  \caption{Parameters of the $M_{\mathrm{200c}}$--$\mu_{\star}$--$z$ relation. Contours show the $1\sigma$, $2\sigma$ and $3\sigma$ confidence areas from the fiducial \texttt{FULL} run. At the top label, we show the $1\sigma$ total uncertainties. For easy comparison with other results in the literature we are plotting $\log_{10} M_0$ converted from $M_{0}$, in which we performed the fit.}
  \label{fig:corner2}
\end{figure}

Our final mass calibration is performed using the masses from our fiducial \verb|Full| configuration together with the mass covariance described in this section. The posteriors of the fitted parameters are summarised in \autoref{tab:mass_richness_parameters}. The corresponding confidence contours are shown in \autoref{fig:corner2}. This result shows that a galaxy cluster with $\mu_{\star}=5.16\times 10^{12} \, {\rm M_{\odot}}$ at $z=0.35$ has a mean mass of $\log_{10} M_{\mathrm{200c}} = 14.06 \pm 0.03$. 

\begin{figure}
  \includegraphics[width=\linewidth]{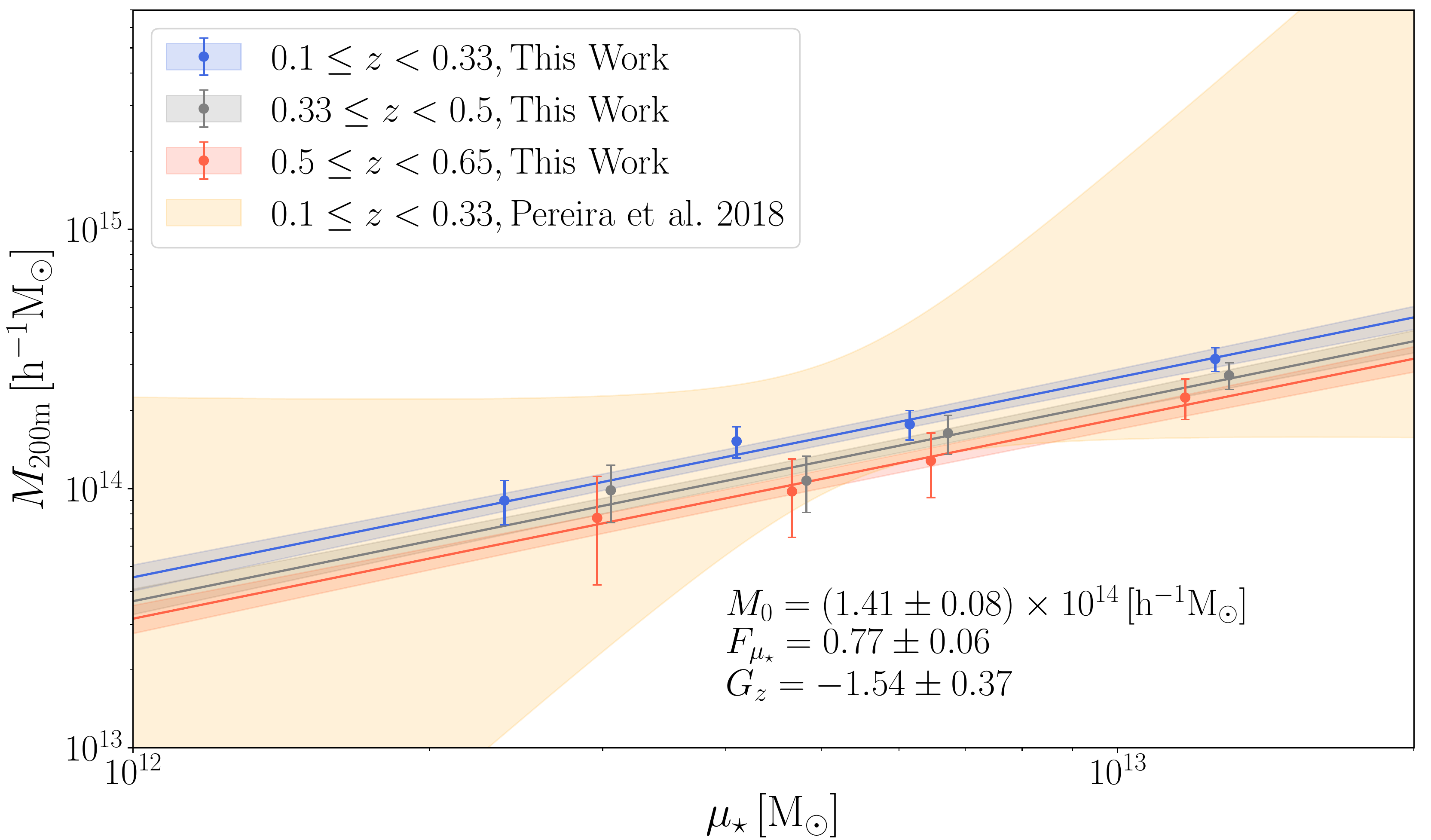}
  \caption{The mass calibration in three redshift bins for the mass definition $M_{\mathrm{200m}}$ in units of ${\rm h^{-1}\, M_{\odot}}$. The \textit{solid} lines are the result of our best-fit with the corresponding $1\sigma$ confidence intervals as \textit{shaded} regions. For comparison, we also present the $2\sigma$ confidence intervals of the previous mass-calibration of \protect\cite{2018MNRAS.474.1361P} as the \textit{orange shaded} region.
  }
  \label{fig:finalmasscalib}
\end{figure}

For a direct comparison with our previous work \citep{2018MNRAS.474.1361P}, we show in \autoref{fig:finalmasscalib} our estimated  $M_{\mathrm{200m}}$--$\mu_{\star}$--$z$ relation (\textit{blue}, \textit{grey} and \textit{red solid} lines) and the corresponding $1\sigma$ confidence intervals (\textit{blue}, \textit{grey} and \textit{red shaded} regions) for redMaPPer clusters in DES Y1 overlapped with the $2\sigma$ confidence intervals (\textit{orange shaded} regions) for the mass-calibration of CS82's redMaPPer clusters.  

\section{Results}
\label{sec:results}

We perform a weak lensing mass calibration of the total stellar-mass based mass proxy $\mu_{\star}$ using DES Y1 redMaPPer clusters. We divide our sample in 12 stacks binned by redshift and $\mu_{\star}$ in the range $0.1 \le z<0.65$, $\mu_{\star} < 5.5 \times 10^{13}\, \mathrm{M_{\odot}}$, using the sample with $\lambda > 20$. Therefore, we use the same redMaPPer-selected cluster sample as in \cite{2019MNRAS.482.1352M}, but compute $\mu_{\star}$ for each of the identified clusters and perform the weak-lensing analysis binning in this new proxy.                    

We model the weak lensing signal by taking into account: cluster miscentring (\autoref{sec:misc_corr}); model calibration systematics (\autoref{sec:mod_sys}); source sample dilution by cluster members (\autoref{sec:mod_boost}); shear measurement systematics and source photometric redshift uncertainties (\autoref{sec:red_shear}, \autoref{sec:shear_photoz}); triaxiality and projection effects (\autoref{sec:triax_proj}). We perform the modelling of the weak lensing signal and apply a blinding factor in the derived posterior of the masses to avoid confirmation bias in our estimates. 

We performed the unblinding after reaching the final version of our modelling pipeline \href{https://github.com/MariaElidaiana/desy1_mustar_analysis}{\faGithubAlt}, which was validated by an internal review from members of the DES collaboration prior to unblinding. No changes to the analysis and modelling pipeline were made post-unblinding.             

We use the derived average masses to determine the cluster mass calibration of $M_{\mathrm{200c}}$ as a function of $\mu_{\star}$ and redshift according to \autoref{eq:mass_richness_redshift}. The summary of our constraints on the scaling relation for clusters at pivots $\mu_{\star}^0=5.16\times 10^{12} \, {\rm M_{\odot}}$ and $z_0=0.35$ is a mean cluster mass of  
\begin{equation}
    M_0 = [1.14 \pm 0.05 \pm 0.05] \cdot 10^{14} \,\mathrm{ h^{-1} M_{\odot}}, 
\label{blinded_M}    
\end{equation}
with the slope $F_{\mu_{\star}}$ for the mass-proxy's term of 
\begin{equation}
    F_{\mu_{\star}} = 0.76 \pm 0.01 \pm 0.06,
\label{slope_F}    
\end{equation}
and the slope $G_{z}$ for the redshift's term of 
\begin{equation}
    G_{z} = -1.14 \pm 0.04 \pm 0.38,
\end{equation}
where the first and second terms in the errors correspond to statistical and systematic, respectively.       

\section{Discussion}
\label{sec:disc}

Here we present a detailed discussion of the relationship between $\mu_{\star}$ and $\lambda$ to check the impact on the MOR of our $\lambda$-selected sample binned in $\mu_{\star}$. We also discuss our findings in the context of our previous work and present some possible ways to use $\mu_{\star}$ for applications in cluster cosmology.

\subsection{The relationship between $\lambda$ and $\mu_{\star}$ and implications for the slope $F_{\mu_{\star}}$}
\label{sec:lamu_relation}

In this section, we describe how to build a model for the relationship between $\lambda$ and $\mu_{\star}$ from simulations to check for potential selection effects. We start with a dark matter halo catalogue from a simulation of sufficient scale to contain rare massive halos. As our work uses a $\lambda\ge 20$ catalogue to bin in $\mu_\star$, we will need to assign each halo $\mu_{\star}$ and $\lambda$.

Following \cite{2014MNRAS.438...78R}, \cite{2014MNRAS.441.3562E}, \cite{2017MNRAS.466.3103S} and \cite{2018PASJ...70S..20O}, we assume the relationship of mass and an observable $s$ follows a power-law form. For $\lambda$, this $\langle \log M|s \rangle$ is
\begin{equation}
    \langle \log M|\lambda \rangle = a \log\left (\frac{\lambda}{40}\right) + b\,, 
\end{equation}
where from \cite{2019MNRAS.482.1352M} we have $a=1.36 \pm 0.05$, $b=14.49 \pm 0.02$ (we will assume redshift evolution is zero), and from \cite{2019MNRAS.490.3341F} we have $\sigma_{\log M|\lambda} = 0.13$. For $\mu_\star$, the relation is
\begin{equation}
    \langle \log M|\mu_{\star} \rangle = a 
    \log \left(\frac{\mu_{\star}}{5.2\times 10^{12}}\right ) + b\,,
\end{equation}
where from this paper we have $a=0.77 \pm 0.06$, $b=14.30\pm 0.02$, and by converting the scatter from \cite{2020MNRAS.493.4591P} to logarithmic base 10, we have that $\sigma_{\log M|\mu_{\star}} = 0.11$. Note that we have made the proper conversions of these results to work with masses from simulations that are in units of $M_{200m}[M_{\odot}]$.   

To assign properties to our dark matter halos we will need $p(\log \mathbf{s}|M)$, where $\mathbf{s}$ is our vector of observables, i. e. $\mathbf{s}=\{\lambda, \,\mu_{\star}\}$. Following \cite{2014MNRAS.441.3562E} and \cite{2018PASJ...70S..20O}, through the Bayes theorem $p(\log M| \mathbf{s})$ can be converted to $p(\log  \mathbf{s}|M)$ by \begin{equation}
p(\log \mathbf{s} |M ) = \frac{p(\log M| \mathbf{s}) p(\log \mathbf{s})}
{\int d(\log \mathbf{s}) p(\log M| \mathbf{s})p(\log \mathbf{s})}. 
\end{equation}
Using the locally power-law model of \cite{2014MNRAS.438...78R} and \cite{2014MNRAS.441.3562E}, we find that this has a log-normal distribution with mean
\begin{equation}
\label{mock_build}
\langle \log \mathbf{s} | M \rangle = \left[ \boldsymbol{\alpha}^T \mathbf{C}^{-1} (\mathbf{\mu} - \boldsymbol{\pi}) - \boldsymbol{\beta} \ln{(10)} \right] \sigma_{\log \mathbf{s}|M}^2,
\end{equation}
and variance 
\begin{equation}
\label{logsM_variance}
\sigma_{\log \mathbf{s}|M}^2 = \left( \boldsymbol{\alpha}^T \mathbf{C}^{-1}  \boldsymbol{\alpha} \right)^{-1},
\end{equation}
where $\boldsymbol{\alpha} =\{a_{\lambda},a_{\mu_\star}\}$, $\mu = \log M$, $\boldsymbol{\pi}=\{b_\lambda, b_{\mu_\star}\}$ and $\beta$ is the slope of the observable's function. For $\lambda$ this is $p(\log \lambda) \propto \lambda^{-\beta}$. Working iteratively, we find that $\beta_\lambda = 3.5$ and $\beta_{\mu_\star} = 1.67$ for mocks near thresholds of interest. The elements of the covariance matrix $\mathbf{C}$ are computed by $C_{ij} = r_{ij}\sigma_i\sigma_j$, where $r$ is the correlation coefficient between $\lambda$ and $\mu_{\star}$. The scatters $\sigma$ are given by the components $\{\sigma_{\log \lambda|M}, \sigma_{\log \mu_\star|M}\}$.  

We took the halo catalogue from the DES Buzzard simulation v1.9.2 \citep{2019arXiv190102401D}, selecting a total of 16,000 square degrees. We paint on the observables accounting for the correlation between them: to a given halo, each observable $\mathbf{s}$ has a property computed via \autoref{mock_build} with a random normal deviation given by \autoref{logsM_variance}. 

The mock catalogue is used to construct observable vectors that may be compared against data or simulations. To verify whether our observable vectors are reliable we can check: $i)$ the observed $p(\lambda|\mu_{\star})$ and $ii)$ the fraction $p(\log \mu_\star-M | \log M)$. We will start with the latter.  
The stellar mass in clusters is known to be a few per cent of the halo dark matter mass. In the simulations studied by \cite{2018MNRAS.478.2618F}, Figure 3 shows a $p(\log M_\star-M | \log M)$ that goes from 2 to 1 per cent over the mass range of interest. We note here that in their work they use the stellar mass $M_{\star}$ and not $\mu_{\star}$, but since we derived $\mu_{\star}$ from stellar mass, we expect to recover similar values for the stellar fraction computed with $\mu_{\star}$, i. e. $p(\log \mu_\star-M | \log M)$. Using \autoref{mock_build}, we can paint stellar masses on top of dark matter halos. In order to recreate a physical behaviour for the $p(\log \mu_\star-M | \log M)$ relation of halos with $\log M/M_\odot \approx 14.30$, the value for the slope needs to be between $F_{\mu_\star}\approx$ 0.75 and 1. We checked that values of $F_{\mu_\star}$ lower than this (e.g. $\sim 0.5$) start to deviate from what is known about the cluster stellar fractions.  

A mock catalogue generated using values for the pivot mass and the mass proxy scaling for $\mu_\star$ that recreates the cluster stellar fractions in the simulations of \cite{2018MNRAS.478.2618F} and the scaling relation of \cite{2019MNRAS.482.1352M} for $\lambda$, then reproduces the observed $p(\lambda|\mu_{\star})$ relation in the log-space, that is
\begin{equation}
    \label{lambda_mu_data}
    \log \lambda = (-5.06 \pm 0.17) + (0.52 \pm 0.01 )\log \mu_\star,
\end{equation}
with intrinsic scatter of $\sigma_{\lambda|\mu_\star} = 0.16$. 
For the mocks, performing a simple linear fit with Python Polyfit, we obtained a relation that is $ \log \lambda = (-6.42\pm0.02) + (0.626\pm 0.002) \log \mu_\star $, with scatter of $\sigma_{\lambda|\mu_\star} = 0.08$. Therefore, when using the fitted intrinsic scatter $\sigma_{\lambda|\mu\star}$ as the $1\sigma$ uncertainty of the mean relation, the derived observed and simulated relations are in good agreement.  

Having produced acceptable mocks we can explore the question of the effect of selecting on $\lambda>20$ on the slope of the $\langle \log M|\mu_\star,z \rangle$ relation measured by weak lensing having stacked in $\mu_\star$. 

In order to do this, we divided the halos in 3 bins of $z$ $[0.1-0.33, 0.33-0.5, 0.5-0.65]$ and 4 of $\mu_\star$ $[0.5-3, 3-5, 5-10, 10-100]\times10^{12}$M$_\odot$ for samples that have cuts of $\lambda \ge [0, 10, 20]$. We took the average values of $M$, $\mu_\star$ and $z$ in these bins and then performed an MCMC fit in the same form we did in real data. We found that the slope between the lowest and second $\lambda$-cut is basically unaffected, changing by $\sim 2$ per cent. The slope between the second and third $\lambda$-cut changed by $\sim 7$ per cent. Considering the error bars from the MCMC, we can see that this change in the slope is not too significant.         
In summary, we believe that selecting on one observable and binning on another should have an effect (e. g. changes in the slope), but looking into simulations that reliably reproduce our observables, we find that there was no significant change in the slope when we mimic this selection. Therefore, we do not believe there is a significant signal of this selection effect in our results. Second, for plausible stellar mass fractions, i. e. $p(\log \mu_\star-M | \log M)$ relations, one expects power-law relations for $\langle \log M|\mu_\star \rangle$ to have exponents between 0.75 and 1, which is consistent with the slope of 0.77 that we found in our data.      

\subsection{The redshift evolution of the $M$--$\mu_{\star}$--$z$ relation }
\label{sec:redevol}

The $M$--$\mu_{\star}$--$z$ relation presented in the Section \ref{sec:results} shows a marginal dependence on redshift. $G_z$ is in fact $3\sigma$ away from the $G_z=0$ case. However, as can be noted from \autoref{fig:finalmasscalib}, there is a $0.1-0.2$ dex difference in $\mu_\star$ at fixed $M_{200m}$ between the lowest and highest redshift bin. This number is consistent with the typical intrinsic scatter in stellar mass at fixed halo mass (e.g. \citealt{pillepich}), implying that the found redshift evolution is not significant. The stellar mass functions in DES galaxy clusters studied in \cite{palmese2020} also find no significant redshift evolution, using the same stellar masses.

Given the current uncertainties, the simple redshift evolution model used in this work is appropriate, but for future analyses with DES clusters including larger statistics, a more sophisticated model shall be tested. In fact, \cite{2018MNRAS.478.2618F} showed that the slope and scatter in the stellar mass--halo mass relation show some evidence for running with $z$, although this is not as strong of an effect as the one found for the gas fraction.

When comparing our result to the literature, one should also note that intra-cluster light (ICL) is not taken into account in this work, since simulation studies will often include the diffuse component when quoting the total stellar mass. The ICL can constitute a significant fraction of the total stellar mass (up to 40 per cent, e.g. \citealt{pillepich} and \citealt{2019ApJ...874..165Z}), and it has been shown to build up since $z\sim 1$ (e.g. \citealt{2015MNRAS.449.2353B}).

\subsection{Comparison with previous work and considerations about selection effects}
\label{sec:seleff}

The comparison of the result obtained in this work with a previous calibration of $\mu_{\star}$ at low redshifts is tricky, because there we used a different cluster sample identified by redMaPPer and Voronoi-Tessellation (VT) cluster finders in the SDSS Stripe 82 region, and we assumed a MOR without redshift evolution. For SDSS redMaPPer clusters, we found a slope for the mass proxy of $1.74 \pm 0.62$, which is compatible at $\sim 2\sigma$ with our present result.

It is known that at $z < 0.1$, nearly all cluster members are red (e.g. \citealt{2007A&A...471...17A}). However, at higher redshifts ($z>0.1$), the number of blue galaxies is observed to increase and the number of red members is observed to decrease (e.g. \citealt{1984ApJ...285..426B, 1995ApJ...439...47R, 2007MNRAS.376.1425G, 2018PASJ...70S..24N}). Furthermore, at low $z$ almost all galaxies more massive than $10^{10.3}\,\mathrm{M_{\odot}}$ are red, and this corresponds roughly to the $0.4\, \mathrm{L_\star}$ luminosity threshold of redMaPPer. Thus, at low $z$, $\lambda$ and $\mu_{\star}$ red-sequence selected samples should have, approximately, the same number of total members and the same stellar mass. At higher $z$, richness and stellar mass are expected to evolve differently. We believe that this effect is related to the evidence for redshift evolution in our MOR results (as the redshift slope $G_z$ is not consistent with zero, see section \ref{sec:redevol}) that is not observed in \cite{2019MNRAS.482.1352M}.  In fact, previous works (e.g. \citealt{2018MNRAS.478.2618F}) have found evidence that the stellar mass content of clusters may evolve with redshift. 

We have checked that the potential selection effect on  $\langle M|\mu_{\star},z \rangle$ introduced by the fact that the cluster sample has been selected with a cut in richness at $\lambda>20$ is subdominant for our results (see \autoref{sec:lamu_relation}). In fact, in the absence of scatter between $\lambda$ and $\mu_{\star}$, selecting in $\lambda$ or in $\mu_{\star}$ would have the same meaning. However, \citet{palmese2020} find that the scatter in $\mu_{\star}$ at fixed richness is $\sigma_{\mu_\star|\lambda}\sim 0.25$ dex for the $\lambda>5$ sample, result which is largely dominated by the scatter at the low-richness end ($\lambda<20$). The largest impact of this scatter on our result is expected to be at the lowest $\mu_{\star}$ binning, where some clusters may have scattered to $\lambda<20$. We tested the impact of this effect by removing from our fit the lowest $\mu_{\star}$ binning, and found no significant change in our parameter estimates of the MOR.

With this work, we complete the program of establishing $\mu_{\star}$ as a reliable mass proxy in the same regime as the $\lambda$-based mass calibration work by the DES collaboration, opening the possibility of exploring the novel regimes of low mass--low $z$ and high mass--high $z$ in a forthcoming paper. 

\subsection{Possible implications for cluster cosmology}
\label{sec:mucosmo}

Since there is a tight connection between galaxy masses and halo masses (e.g. \citealt{2009ApJ...696..620C, 2010ApJ...717..379B, 2015MNRAS.449.1352C, 2017ApJ...840..104S, 2017MNRAS.471.1153N, 2018ARA&A..56..435W, 2020MNRAS.492.3685H, palmese2020}), stellar-mass based mass proxies such as $\mu_{\star}$ show great potential to be accurate halo mass estimators in photometric galaxy surveys. They can be used to probe galaxy evolution and can also help to improve the constraints on cosmological parameters. 

In the review by \cite{2018ARA&A..56..435W}, they present a series of application for the galaxy-halo connection in cosmology, e. g. systematics in cluster cosmology, the impact of baryons and galaxy clustering at small scales. For cluster cosmology, in particular, several studies have shown that projection effects have a non-negligible impact on the mass-richness relationship \citep{2018MNRAS.481..324W, 2019MNRAS.482..490C, 2019PASJ...71..107M, 2020arXiv200203867S}, most likely due to a dependence on the details of the galaxy-halo connection, such as the colour dependence of the cluster members. Since $\mu_{\star}$  is a colour-independent proxy and has a well-defined physical interpretation, we believe it has the potential to contribute in the understanding of the projection effects in the cluster cosmology context. In a future work, we plan to perform a comparison of projection effects between $\mu_{\star}$ and $\lambda$, using the new version of the DES Buzzard simulation \citep{2019arXiv190102401D} that has stellar-mass information. 

\section{Summary}
\label{sec:summ}

We have measured the stacked weak lensing signal around 6,124 clusters in the DES Y1 redMaPPer catalogue with $\lambda>20$ and $0.1 \leq z<0.65$. We have computed the stellar-mass based proxy $\mu_{\star}$ for these clusters and performed the lensing measurements in bins of $\mu_{\star}$ and $z$. In the mass modelling, we have accounted for several systematics including cluster miscentring, model calibration, boost-factors, shear and photo-z bias, triaxiality and projection effects.             

Then, we use the fitted weak-lensing mass to perform the mass calibration of this sample. We find a mass--$\mu_{\star}$--$z$ relation of  
\begin{equation}
\begin{aligned}
\langle M_{\mathrm{200c}}| \mu_{\star},z \rangle = &1.14 \pm 0.05 \, \mathrm{stat.} \pm 0.05 \, \mathrm{sys.} \cdot 10^{14}  \\
                                   &\times \, \left( \frac{\mu_{\star}}{5.16\times 10^{12}\mathrm{M_{\sun}}} \right)^{ 0.76 \pm 0.01\, \mathrm{stat.} \pm 0.06 \, \mathrm{sys.} } \\ 
                                   &\times \left(\frac{1+z}{1.35}\right)^{-1.14 \pm 0.04 \, \mathrm{stat.} \pm 0.38 \, \mathrm{sys.} },  
\end{aligned}
\end{equation}
in units of $h^{-1} \mathrm{M_{\sun}}$. This scaling relation is consistent within $2\sigma$ with previous $\mu_{\star}$ measurements using the SDSS redMaPPer clusters and lensing data from CS82 survey \citep{2018MNRAS.474.1361P}.

We have used mock catalogues from DES Buzzard simulations to check for a signal of selection effects since we have a $\lambda$--selected sample binned in $\mu_{\star}$, but we found that such signal is negligible. We also concluded that if such an effect is present, we should have seen a considerable change in the slope due to the lowest $\mu_{\star}$ bin in comparison to the other bins. We test this hypothesis in the data by removing the lowest bin of $\mu_{\star}$ and performing the mass-calibration again. We found no significant change in the slope of our relation. Therefore, we conclude that our analysis is not significantly affected by this selection effect. However, we understand that further work to properly quantify this selection effect is necessary. We also show from these mocks that shallower slopes in the mass proxy term are possible for stellar-mass based proxies. 

We found evidence for redshift evolution in our scaling relation. However, the difference in $\mu_{\star}$ at fixed halo mass between the lowest and highest redshift bin is $\sim 0.1-0.2$ dex, which is consistent with intrinsic scatter in stellar mass at fixed halo mass, and this implies that the redshift evolution we found might not be significant.    

This work provides the most careful weak-lensing mass calibration of $\mu_{\star}$ to date. It is an important step towards establishing $\mu_{\star}$ as a reliable mass proxy not only for studying systematics such as projection effects and low richness clusters but also for future applications in cluster cosmology.    

\section*{Acknowledgements}

This paper has been internally evaluated against a set of pass/fail criteria to give us confidence in its correctness prior to unblinding. 

MESP thanks Ben Lillard and Felix Kling for comments on the manuscript.

Funding for the DES Projects has been provided by the U.S. Department of Energy, the U.S. National Science Foundation, the Ministry of Science and Education of Spain, 
the Science and Technology Facilities Council of the United Kingdom, the Higher Education Funding Council for England, the National Center for Supercomputing 
Applications at the University of Illinois at Urbana-Champaign, the Kavli Institute of Cosmological Physics at the University of Chicago, 
the Center for Cosmology and Astro-Particle Physics at the Ohio State University,
the Mitchell Institute for Fundamental Physics and Astronomy at Texas A\&M University, Financiadora de Estudos e Projetos, 
Funda{\c c}{\~a}o Carlos Chagas Filho de Amparo {\`a} Pesquisa do Estado do Rio de Janeiro, Conselho Nacional de Desenvolvimento Cient{\'i}fico e Tecnol{\'o}gico and 
the Minist{\'e}rio da Ci{\^e}ncia, Tecnologia e Inova{\c c}{\~a}o, the Deutsche Forschungsgemeinschaft and the Collaborating Institutions in the Dark Energy Survey. 

The Collaborating Institutions are Argonne National Laboratory, the University of California at Santa Cruz, the University of Cambridge, Centro de Investigaciones Energ{\'e}ticas, 
Medioambientales y Tecnol{\'o}gicas-Madrid, the University of Chicago, University College London, the DES-Brazil Consortium, the University of Edinburgh, 
the Eidgen{\"o}ssische Technische Hochschule (ETH) Z{\"u}rich, 
Fermi National Accelerator Laboratory, the University of Illinois at Urbana-Champaign, the Institut de Ci{\`e}ncies de l'Espai (IEEC/CSIC), 
the Institut de F{\'i}sica d'Altes Energies, Lawrence Berkeley National Laboratory, the Ludwig-Maximilians Universit{\"a}t M{\"u}nchen and the associated Excellence Cluster Universe, 
the University of Michigan, the National Optical Astronomy Observatory, the University of Nottingham, The Ohio State University, the University of Pennsylvania, the University of Portsmouth, 
SLAC National Accelerator Laboratory, Stanford University, the University of Sussex, Texas A\&M University, and the OzDES Membership Consortium.

Based in part on observations at Cerro Tololo Inter-American Observatory, National Optical Astronomy Observatory, which is operated by the Association of 
Universities for Research in Astronomy (AURA) under a cooperative agreement with the National Science Foundation.

The DES data management system is supported by the National Science Foundation under Grant Numbers AST-1138766 and AST-1536171.
The DES participants from Spanish institutions are partially supported by MINECO under grants AYA2015-71825, ESP2015-66861, FPA2015-68048, SEV-2016-0588, SEV-2016-0597, and MDM-2015-0509, 
some of which include ERDF funds from the European Union. IFAE is partially funded by the CERCA program of the Generalitat de Catalunya.
Research leading to these results has received funding from the European Research
Council under the European Union's Seventh Framework Program (FP7/2007-2013) including ERC grant agreements 240672, 291329, and 306478.
We  acknowledge support from the Brazilian Instituto Nacional de Ci\^encia
e Tecnologia (INCT) e-Universe (CNPq grant 465376/2014-2).

This manuscript has been authored by Fermi Research Alliance, LLC under Contract No. DE-AC02-07CH11359 with the U.S. Department of Energy, Office of Science, Office of High Energy Physics.





\bibliographystyle{mnras}
\bibliography{mnras} 




\appendix

\section{Affiliations}

$^{1}$ Brandeis University, Physics Department, 415 South Street, Waltham MA 02453\\
$^{2}$ Fermi National Accelerator Laboratory, P. O. Box 500, Batavia, IL 60510, USA\\
$^{3}$ Kavli Institute for Cosmological Physics, University of Chicago, Chicago, IL 60637, USA\\
$^{4}$ Max Planck Institute for Extraterrestrial Physics, Giessenbachstrasse, 85748 Garching, Germany\\
$^{5}$ Universit\"ats-Sternwarte, Fakult\"at f\"ur Physik, Ludwig-Maximilians Universit\"at M\"unchen, Scheinerstr. 1, 81679 M\"unchen, Germany\\
$^{6}$ Department of Physics, University of Arizona, Tucson, AZ 85721, USA\\
$^{7}$ Department of Astronomy, The Ohio State University, Columbus, OH 43210, USA\\
$^{8}$ Department of Physics, University of Michigan, Ann Arbor, MI 48109, USA\\
$^{9}$ Center for Cosmology and Astro-Particle Physics, The Ohio State University, Columbus, OH 43210, USA\\
$^{10}$ Department of Astronomy, University of California, Berkeley,  501 Campbell Hall, Berkeley, CA 94720, USA\\
$^{11}$ Santa Cruz Institute for Particle Physics, Santa Cruz, CA 95064, USA\\
$^{12}$ Institut de F\'{\i}sica d'Altes Energies (IFAE), The Barcelona Institute of Science and Technology, Campus UAB, 08193 Bellaterra (Barcelona) Spain\\
$^{13}$ Department of Physics, Stanford University, 382 Via Pueblo Mall, Stanford, CA 94305, USA\\
$^{14}$ Kavli Institute for Particle Astrophysics \& Cosmology, P. O. Box 2450, Stanford University, Stanford, CA 94305, USA\\
$^{15}$ SLAC National Accelerator Laboratory, Menlo Park, CA 94025, USA\\
$^{16}$ D\'{e}partement de Physique Th\'{e}orique and Center for Astroparticle Physics, Universit\'{e} de Gen\`{e}ve, 24 quai Ernest Ansermet, CH-1211 Geneva, Switzerland\\
$^{17}$ Department of Physics \& Astronomy, University College London, Gower Street, London, WC1E 6BT, UK\\
$^{18}$ Department of Physics, ETH Zurich, Wolfgang-Pauli-Strasse 16, CH-8093 Zurich, Switzerland\\
$^{19}$ Department of Physics, The Ohio State University, Columbus, OH 43210, USA\\
$^{20}$ Department of Physics and Astronomy, University of Pennsylvania, Philadelphia, PA 19104, USA\\
$^{21}$ Department of Physics and Astronomy, Stony Brook University, Stony Brook, NY 11794, USA\\
$^{22}$ Institute for Astronomy, University of Edinburgh, Edinburgh EH9 3HJ, UK\\
$^{23}$ Cerro Tololo Inter-American Observatory, NSF's National Optical-Infrared Astronomy Research Laboratory, Casilla 603, La Serena, Chile\\
$^{24}$ Departamento de F\'isica Matem\'atica, Instituto de F\'isica, Universidade de S\~ao Paulo, CP 66318, S\~ao Paulo, SP, 05314-970, Brazil\\
$^{25}$ Laborat\'orio Interinstitucional de e-Astronomia - LIneA, Rua Gal. Jos\'e Cristino 77, Rio de Janeiro, RJ - 20921-400, Brazil\\
$^{26}$ Instituto de Fisica Teorica UAM/CSIC, Universidad Autonoma de Madrid, 28049 Madrid, Spain\\
$^{27}$ CNRS, UMR 7095, Institut d'Astrophysique de Paris, F-75014, Paris, France\\
$^{28}$ Sorbonne Universit\'es, UPMC Univ Paris 06, UMR 7095, Institut d'Astrophysique de Paris, F-75014, Paris, France\\
$^{29}$ Department of Physics and Astronomy, Pevensey Building, University of Sussex, Brighton, BN1 9QH, UK\\
$^{30}$ Jodrell Bank Center for Astrophysics, School of Physics and Astronomy, University of Manchester, Oxford Road, Manchester, M13 9PL, UK\\
$^{31}$ Instituto de Astrofisica de Canarias, E-38205 La Laguna, Tenerife, Spain\\
$^{32}$ Universidad de La Laguna, Dpto. Astrofsica, E-38206 La Laguna, Tenerife, Spain\\
$^{33}$ Department of Astronomy, University of Illinois at Urbana-Champaign, 1002 W. Green Street, Urbana, IL 61801, USA\\
$^{34}$ National Center for Supercomputing Applications, 1205 West Clark St., Urbana, IL 61801, USA\\
$^{35}$ INAF-Osservatorio Astronomico di Trieste, via G. B. Tiepolo 11, I-34143 Trieste, Italy\\
$^{36}$ Institute for Fundamental Physics of the Universe, Via Beirut 2, 34014 Trieste, Italy\\
$^{37}$ Observat\'orio Nacional, Rua Gal. Jos\'e Cristino 77, Rio de Janeiro, RJ - 20921-400, Brazil\\
$^{38}$ Department of Physics, IIT Hyderabad, Kandi, Telangana 502285, India\\
$^{39}$ Faculty of Physics, Ludwig-Maximilians-Universit\"at, Scheinerstr. 1, 81679 Munich, Germany\\
$^{40}$ Institut d'Estudis Espacials de Catalunya (IEEC), 08034 Barcelona, Spain\\
$^{41}$ Institute of Space Sciences (ICE, CSIC),  Campus UAB, Carrer de Can Magrans, s/n,  08193 Barcelona, Spain\\
$^{42}$ Department of Astronomy, University of Michigan, Ann Arbor, MI 48109, USA\\
$^{43}$ School of Mathematics and Physics, University of Queensland,  Brisbane, QLD 4072, Australia\\
$^{44}$ Center for Astrophysics $\vert$ Harvard \& Smithsonian, 60 Garden Street, Cambridge, MA 02138, USA\\
$^{45}$ Australian Astronomical Optics, Macquarie University, North Ryde, NSW 2113, Australia\\
$^{46}$ Lowell Observatory, 1400 Mars Hill Rd, Flagstaff, AZ 86001, USA\\
$^{47}$ George P. and Cynthia Woods Mitchell Institute for Fundamental Physics and Astronomy, and Department of Physics and Astronomy, Texas A\&M University, College Station, TX 77843,  USA\\
$^{48}$ Department of Astrophysical Sciences, Princeton University, Peyton Hall, Princeton, NJ 08544, USA\\
$^{49}$ Instituci\'o Catalana de Recerca i Estudis Avan\c{c}ats, E-08010 Barcelona, Spain\\
$^{50}$ Institute of Astronomy, University of Cambridge, Madingley Road, Cambridge CB3 0HA, UK\\
$^{51}$ Centro de Investigaciones Energ\'eticas, Medioambientales y Tecnol\'ogicas (CIEMAT), Madrid, Spain\\
$^{52}$ School of Physics and Astronomy, University of Southampton,  Southampton, SO17 1BJ, UK\\
$^{53}$ Computer Science and Mathematics Division, Oak Ridge National Laboratory, Oak Ridge, TN 37831\\



\bsp	
\label{lastpage}
\end{document}